\RequirePackage{fix-cm}
\documentclass[twocolumn,epjc3]{svjour3}  
\smartqed  
\RequirePackage{graphicx}
\RequirePackage[colorlinks,citecolor=blue,urlcolor=blue,linkcolor=blue]{hyperref} 
\usepackage{color}
\usepackage{ifthen}
\newboolean{makefigures}                
\setboolean{makefigures}{true}          
\usepackage{upgreek}
\journalname{Eur. Phys. J. C}
\newcommand{\ctsper}      {cts/(keV$\cdot$kg$\cdot$yr)}

\newcommand{\zctsper}     {{$10^{-2}$~cts/(keV$\cdot$kg$\cdot$yr)}}

\newcommand{\dctsper}     {{$10^{-3}$~cts/(keV$\cdot$kg$\cdot$yr)}}

\newcommand{\vctsper}     {{$10^{-4}$~cts/(keV$\cdot$kg$\cdot$yr)}}

\newcommand{\kgy}         {{kg$\cdot$yr}}
\newcommand{\kgyr}        {{kg$\cdot$yr}}

\newcommand{\cum}         {{m$^3$}}
\newcommand{\mubq}        {{$\upmu$Bq}}
\newcommand{\mum}         {{$\upmu$m}}
\newcommand{\mus}         {{$\upmu$s}}
\newcommand{\mubqperkg}   {{${\upmu\mathrm{Bq}}/{\mathrm{kg}}$}}
\def\cpowten#1#2{{$#1\cdot10^{#2}$}}
\def\powten#1{{$10^{#1}$}}

\newcommand{\qbb}         {{$Q_{\beta\beta}$}}

\newcommand{\onbb}        {{$0\nu\beta\beta$}}
\newcommand{\nnbb}        {{$2\nu\beta\beta$}}

\newcommand{\up}          {\rule{0mm}{5mm}}

\newcommand{\etal}        {\textit{et al.}}

\newcommand{\gerda}       {\textsc{Gerda}}

\newcommand{\GERDA}       {\mbox{\textsc{Gerda}}}  
\newcommand{\LNGS}        {{\mbox{\textsc{Lngs}}}}
\newcommand{\WT}          {water tank}

\newcommand{\GeMPI}       {Ge\textsc{MPI}}

\newcommand{\majorana}    {\textsc{Majorana}}

\newcommand{\igex}        {\textsc{Igex}}

\newcommand{\hdm}         {\textsc{HdM}}

\newcommand{\borexino}       {\mbox{\textsc{Borexino}}}

\newcommand{\Gallex}      {\textsc{Gallex}}
\newcommand{\LVD}         {{\mbox{\textsc{Lvd}}}}

\newcommand{\geant}       {\textsc{Geant4}}

\newcommand{\rootv}       {\textsc{Root}}

\newcommand{\mage}        {\textsc{MaGe}}
\newcommand{\gelatio}     {\textsc{Gelatio}}
\newcommand{\mgdo}        {\mbox{MGDO}}
\newcommand{\tier}        {\textsc{Tier}}
\newcommand{\gesix}       {{$^{76}$Ge}}

\newcommand{\geenr}       {{$^{\rm enr}$Ge}}          
\newcommand{\genat}       {{$^{\rm nat}$Ge}}
\newcommand{\gedep}       {{$^{\rm dep}$Ge}}

\newcommand{\thzza}       {{$^{228}$Th}}

\newcommand{\Rn}          {$^{222}$Rn}
\newcommand{\Ra}          {$^{226}$Ra}

\newcommand{\Th}          {$^{228}$Th}

\newcommand{\Co}          {$^{60}$Co}

\begin{document}

\title{The \mbox{\sc{Gerda}} experiment for the search of 
   $0\nu\beta\beta$  decay in $^{76}$Ge
}

\titlerunning{The \textsc{Gerda} experiment}   

\author{
   K.-H.~Ackermann\thanksref{MPIP} \and                
   M.~Agostini\thanksref{TUM} \and
   M.~Allardt\thanksref{DD} \and
   M.~Altmann\thanksref{MPIP,deceased} \and
   E.~Andreotti\thanksref{GEEL,TU} \and
   A.M.~Bakalyarov\thanksref{KU} \and
   M.~Balata\thanksref{ALNGS} \and
   I.~Barabanov\thanksref{INR} \and
   M.~Barnab\'e Heider\thanksref{HD,TUM,nowCAN} \and              
   N.~Barros\thanksref{DD}
% ...
}  

\authorrunning{the \textsc{Gerda} collaboration}

\thankstext{deceased}{deceased}
\thankstext{nowCAN}{\emph{Present Address:} CEGEP St-Hyacinthe,
 Qu{\'e}bec, Canada}
\thankstext{nowSudbury}{\emph{Present Address:} Laurentian University,
  Sudbury, Canada  }
\thankstext{nowBNL}{\emph{Present Address:} Brookhaven National Laboratory,
    Upton (NY), USA}
\thankstext{nowCHIE}{\emph{Present Address:} Department of Neurosciences and
  Imaging, University ``G. d’Annunzio'' di Chieti-Pescara, Italy}
\thankstext{nowNPL}{\emph{Present Address:} Nat. Physical Laboratory,
  Teddigton, UK}
\thankstext{private}{now in private business}
\thankstext{MIPT}{\emph{Present Address:} Moscow Institute of Physics and
  Technology, Russia} 
\thankstext{nowGOE}{\emph{Present Address:} II. Physikalisches Institut, U. 
       G{\"o}ttingen, Germany, and Department Physik, U. Siegen, Germany}
\thankstext{nowTOKYO}{\emph{Present Address:} Kavli IPMU, University of Tokyo,
  Japan}
\thankstext{nowTUEV}{\emph{Present Address:} T{\"U}V-S{\"U}D, M{\"u}nchen,
  Germany}
\thankstext{nowKIT}{\emph{Present Address:} Karlsruhe Institute of Technology
  (KIT), Karlsruhe, Germany}
\thankstext{nowBOSCH}{\emph{Present Address:} Robert Bosch GmbH, Reutlingen,
  Germany} 
\thankstext{nowDESY}{\emph{Present Address:} Photon-Science Detector Group,
       DESY}
\thankstext{AQU}{\emph{Present Address:} University of L'Aquila, Dipartimento
        di Fisica, L'Aquila, Italy}
\thankstext{corrauthor}{\emph{Corresponding Author},
                                email: grabmayr@uni-tuebingen.de}
\institute{
INFN Laboratori Nazionali del Gran Sasso, LNGS, Assergi, Italy\label{ALNGS} \and
Institute of Physics, Jagiellonian University, Cracow, Poland\label{CR} \and
Institut f{\"u}r Kern- und Teilchenphysik, Technische Universit{\"a}t Dresden,
      Dresden, Germany\label{DD} \and
Joint Institute for Nuclear Research, Dubna, Russia\label{JINR} \and
Institute for Reference Materials and Measurements, Geel,
     Belgium\label{GEEL} \and
Max Planck Institut f{\"u}r Kernphysik, Heidelberg, Germany\label{HD} \and
Dipartimento di Fisica, Universit{\`a} Milano Bicocca,
     Milano, Italy\label{MIBF} \and
INFN Milano Bicocca, Milano, Italy\label{MIBINFN} \and
Dipartimento di Fisica, Universit{\`a} degli Studi di Milano e INFN Milano,
    Milano, Italy\label{MILUINFN} \and
Institute for Nuclear Research of the Russian Academy of Sciences,
    Moscow, Russia\label{INR} \and
Institute for Theoretical and Experimental Physics,
    Moscow, Russia\label{ITEP} \and
National Research Centre ``Kurchatov Institute'', Moscow, Russia\label{KU} \and
Max-Planck-Institut f{\"ur} Physik, M{\"u}nchen, Germany\label{MPIP} \and
Physik Department and Excellence Cluster Universe,
    Technische  Universit{\"a}t M{\"u}nchen, Germany\label{TUM} \and
Dipartimento di Fisica e Astronomia dell{`}Universit{\`a} di Padova,
    Padova, Italy\label{PDUNI} \and
INFN  Padova, Padova, Italy\label{PDINFN} \and
Shanghai Jiaotong University, Shanghai, China\label{SJU} \and
Physikalisches Institut, Eberhard Karls Universit{\"a}t T{\"u}bingen,
    T{\"u}bingen, Germany\label{TU} \and
Physik Institut der Universit{\"a}t Z{\"u}rich, Z{\"u}rich,
    Switzerland\label{UZH}
}

\date{Received: date / Accepted: date}

\maketitle

\begin{abstract}
       The \GERDA\ collaboration is performing a search for neutrinoless
       double beta decay of \gesix\ with the eponymous detector.  The
       experiment has been installed and commissioned at the Laboratori
       Nazionali del Gran Sasso and has started operation in November 2011.
       The design, construction and first operational results are described,
       along with detailed information from the R\&D phase.
\end{abstract}
\section{Introduction}
 \label{sec:intro}

 The \gerda\ experiment (GERmanium Detector Array~\cite{Gerda}) is a search
 for the neutrinoless double beta (\onbb) decay of $^{76}$Ge. The observation
 of such a decay would prove that lepton number is not conserved, and that the
 neutrino has a Majorana component~\cite{schechtervalle}.  A discovery of
 \onbb\ decay would have significant implications on particle physics and
 other fields, including cosmology~\cite{cosmo}. The importance of the topic
 has stimulated the development of several experimental approaches to the
 search for \onbb\ decay on a number of isotopes which undergo double beta
 decay. For recent reviews on the state of knowledge concerning double beta
 decay and on running or planned experiments, see
 Refs.~\cite{avignone,reviewonbb1,reviewonbb2,reviewonbb3,reviewonbb4}.

 The experimental signature for \onbb\ decay is a line in the summed electron
 energy spectrum appearing at the $Q$-value for the reaction, \qbb.  The
 experimental result is a measurement of, or a limit on, the half life,
 $T_{1/2}$, for the process. Within the three neutrino model and assuming the
 existence of a significant Majorana component a positive observation of
 \onbb\ decay would possibly give access to the neutrino mass hierarchy as
 well as information on the absolute values of the neutrino masses.  The
 latter is only possible with knowledge of the nuclear matrix elements, ${\cal
   M}^{0\nu}$, as discussed in Ref.~\cite{doi,suhonen,poves,rodin,barea}.  The
 statements on the mass also require an understanding of whether the
 \onbb\ process is solely due to the Majorana nature of the neutrino, or
 whether additional new physics processes beyond the Standard Model
 contribute. A recent review of the particle physics implications of a
 discovery of \onbb\ decay was given in Ref.~\cite{rodejohann}.

 Nuclides that are potentially \onbb\ emitters will decay via the Standard
 Model allowed \nnbb\ decay.  Both reactions are second order weak
 interactions, and therefore have extremely long half lives.  Values have been
 directly measured for \nnbb\ decay in about ten cases and these are in the
 range 10$^{19}$--10$^{21}$ yr~\cite{reviewonbb1}. The half lives for
 \onbb\ decay, assuming the process exists, are expected to be substantially
 longer. Consequently, \onbb\ decay experiments must be sensitive to just a
 few events per year for a source with a mass of tens to hundreds of
 kilograms. Backgrounds must typically be reduced to the level of one event
 per year in the region of interest (ROI), an energy interval of the order of
 the energy resolution around \qbb.
 
 Experiments looking for \onbb\ decay of \gesix\ operate germanium diodes
 normally made from enriched material, i.e.~the number of \gesix\ nuclei, the
 isotopic fraction $f_{76}$, is enlarged from 7.8~\% to 86~\% or higher. In
 these type of experiments, the source is equal to the detector which yields
 high detection efficiency.  Additional advantages of this technique are the
 superior energy resolution of 0.2~\% at \qbb=2039~keV compared to other
 searches with different isotopes and the high radiopurity of the crystal
 growing procedure. Disadvantages are the relatively low \qbb\ value since
 backgrounds typically fall with energy and the relative difficulty to scale
 to larger mass compared to e.g.~experiments using liquids and gases.  There
 is a considerable history to the use of $^{76}$Ge for the search for
 \onbb\ decay.  After initial experiments~\cite{oldge}, the Heidelberg-Moscow
 (\hdm) collaboration~\cite{hdm} and \igex~\cite{igex} were the driving forces
 in this field setting the most stringent limits. In 2004 a subgroup of the
 \hdm\ collaboration~\cite{hvkkclaim} claimed a 4$\sigma$ significance for the
 observation of \onbb\ decay with a best value of
 $T_{1/2}$=1.19$\cdot10^{25}$~yr; the quoted 3\,$\sigma$ range is
 $(0.69-4.19)\cdot10^{25}$ yr.  To scrutinize this result, and to push the
 sensitivity to much higher levels, two new $^{76}$Ge experiments have been
 initiated: \majorana~\cite{majorana} and \gerda~\cite{Gerda}. The latter has
 been built in the INFN Laboratori Nazionali del Gran Sasso (LNGS) at a depth
 of 3500\,m\,w.e. (water equivalent).  Whereas \majorana\ further refines the
 background reduction techniques in the traditional approach of operating
 germanium detectors in vacuum, \gerda\ submerses bare high-purity germanium
 detectors enriched in $^{76}$Ge into liquid argon (LAr) following a
 suggestion by Ref.~\cite{heusser}; LAr serves simultaneously as a shield
 against external radioactivity and as cooling medium. Phase~I of the
 experiment is currently taking data and will continue until a sensitivity is
 reached which corresponds to an exposure of 15~\kgy\ with a background index
 (BI) of 10$^{-2}$~\ctsper~\cite{Gerda}).  This will be sufficient to make a
 strong statement on the existence of \onbb\ decay in $^{76}$Ge for the best
 value given in Ref.~\cite{hvkkclaim}. Phase~II of \gerda\ is planned to
 acquire an exposure of 100~\kgy\ at a BI of 10$^{-3}$~\ctsper.  For pure
 Majorana exchange and the case that no signal is seen, this will constrain
 the effective neutrino mass $\langle m_{\beta\beta}\rangle$ to less than
 about $100$~meV with the precise value depending on the choice of matrix
 elements~\cite{nme}.

 The \gerda\ experiment is described in detail in the following sections. An
 overview of experimental constraints and the design is presented first.  This
 is followed by a description of the Ge detectors.  Then, the experimental
 setup, electronic readout, data acquisition (DAQ) and data processing are
 described. As \gerda\ Phase~I has been fully commissioned and has started data
 production, the main characteristics of its performance are given in the
 final section.

\section{Design and general layout}
\label{sec:exp}

 The experimental challenge is to have nearly background free conditions in
 the ROI around \qbb.  Typically, background levels are quoted in units of
 counts per keV per kilogram per year, \ctsper, since the number of background
 events roughly scales with the detector mass, energy resolution and running
 time.  Defining $\Delta$ as the width of the ROI where a signal is searched
 for, the expected background is the BI multiplied by $\Delta$ in keV and the
 exposure in \kgyr.  \gerda\ has set the goal to keep the expected background
 below 1~event.  For $\Delta=5$~keV and exposures mentioned above, this
 implies a BI of 0.01 and 0.001~\ctsper, respectively, for the two phases of
 \gerda.

 The main feature of the \gerda\ design is to operate bare Ge detectors made
 out of material enriched in $^{76}$Ge (\geenr) in LAr.  This design concept
 evolved from a proposal to operate Ge detectors in liquid nitrogen (LN$_2$)
 \cite{heusser}. It allows for a significant reduction in the cladding
 material around the diodes and the accompanying radiation sources as compared
 to traditional Ge experiments.  Furthermore, the background produced by
 interactions of cosmic rays is lower than for the traditional concepts of
 \hdm, \igex\ or \majorana\ due to the lower Z of the shielding
 material. Other background sources include neutrons and gammas from the
 decays in the rock of the underground laboratory, radioactivity in support
 materials, radioactive elements in the cryogenic liquid (intrinsic, such as
 $^{39}$Ar and $^{42}$Ar, as well as externally introduced, such as radon) as
 well as internal backgrounds in the Ge diodes. These backgrounds were
 considered in the design and construction phase of \gerda\, and resulted in
 specific design choices, selection of materials used and also in how
 detectors were handled.

 Natural Ge (\genat) contains about 7.8\% $^{76}$Ge, and could in
 principle be used directly for a \onbb\ decay experiment. Indeed, the first
 searches for \onbb\ decay used natural Ge detectors~\cite{oldge}.  Enriched
 detectors allow for a better signal-to-background ratio and also yield
 reduced costs for a fixed mass of $^{76}$Ge in the experiment.  The
 improvement in signal-to-background ratio originates from two sources: ($i$)
 many background sources, such as backgrounds from external gamma rays, are
 expected to scale with the total mass of the detector; and ($ii$) the
 cosmogenic production of $^{68}$Ge and $^{60}$Co in the Ge diodes occurs at a
 higher rate for \genat\ than for \geenr.  The lower overall cost is due to
 the fact that the high cost of enrichment is more than offset by the cost of
 producing the extra crystals and diodes required for \genat\ detectors.

\begin{figure}[t]
\begin{center}
\ifmakefigures%
 \includegraphics[width=\columnwidth]{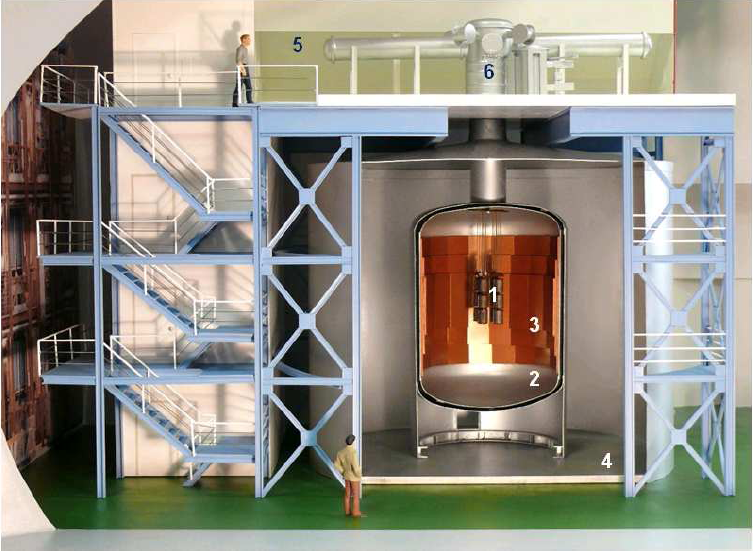}
\fi%
  \caption{ \label{fig:overview}
Artists view (Ge array not to scale) of the \gerda\ experiment as described in
detail in the following sections:
 the germanium detector array (1), 
 the LAr cryostat (2)  with
 its internal copper shield (3) and
 the surrounding water tank (4)  housing the Cherenkov
 muon veto,
 the \gerda\ building with the superstructure supporting 
 the clean room (5)  and
 the lock (6, design modified).
 Various laboratories behind the staircase include the water plant and a radon
 monitor, control rooms, cryogenic infrastructure and the electronics for the
 muon veto.
}
\end{center}
\end{figure}

 Fig.~\ref{fig:overview} shows a model of the realized design: the core of the
 experiment is an array of germanium diodes suspended in strings into a
 cryostat filled with LAr.  The LAr serves both as cooling medium and shield.
 The cryostat is a steel vessel with a copper lining used primarily to reduce
 the gamma radiation from the steel vessel.  The cryostat is placed in a large
 water tank, that fulfills the functions of shielding the inner volumes from
 radiation sources within the hall, such as neutrons, as well as providing a
 sensitive medium for a muon veto system.  A similar experimental setup has
 been proposed previously in Ref.~\cite{zdesenko}.  The detectors are lowered
 into the LAr volume using a lock system located in a clean room on top of the
 water tank.  A further muon veto system is placed on top of the clean room in
 order to shield the neck region of the cryostat.  These installations are
 supported by a steel superstructure.  All components are described in the
 subsequent sections.

\subsection{Auxiliary installations}
  \label{ssec:gdl}
 
 The \gerda\ detector laboratory (GDL), located underground at LNGS, has been
 used for R\&D for \gerda\ as well as for auxiliary tests. It is a grey room
 equipped with a clean bench, a glove box and wet chemistry for etching.  Ge
 diodes submerged in LN$_2$ or LAr can be characterized in a clean environment
 without any exposure to air. The detector handling described in
 sec.~\ref{sec:det} and now adopted for \gerda\ was developed in GDL.
  
 The Liquid Argon Germanium (\textsc{LArGe}) apparatus was installed inside
 GDL to investigate properties of LAr, such as the scintillation light output.
 It is used for studies of background suppression in germanium detectors by
 observing the coincident scintillation light of the liquid
 argon~\cite{DiMarco:2007mr} and to exploit the LAr scintillation light pulse
 shape properties to recognize the interacting particle~\cite{Peiffer:2008zz}.
 \textsc{LArGe} is a 1~m$^3$ low-background copper cryostat with a shield
 consisting of (from outside to inside) 20~cm polyethylene, 23~cm steel, 10~cm
 lead and 15~cm copper. The inner walls are covered with a reflector foil with
 a wavelength shifter coating. The shifted light is detected by nine 8''
 ETL~9357 photomultiplier tubes (PMTs) from Electron Tubes Limited
 (ETL)~\cite{etl}. Calibration sources ($^{228}$Th, $^{226}$Ra, $^{60}$Co,
 $^{137}$Cs) have been placed in- and outside of the cryostat and the event
 rejection by pulse shape discrimination and scintillation light detection
 were studied~\cite{large}.  As a consequence of these measurements
 \gerda\ decided to implement a LAr scintillation light veto for Phase~II.
 \textsc{LArGe} has also been used to understand the background coming from
 the decay of $^{42}$Ar.
 
 In addition to GDL, screening facilities at LNGS, in particular
 GeMPI~\cite{gempi} and Gator~\cite{gator}, have been used extensively.
 Additional screening facilities have been used at different locations,
 including Heidelberg, Geel, and Baksan.
 
 Finally, many of the institutes in the \gerda\ collaboration have
 laboratories which have been extensively used in R\&D and testing related to
 the experiment.

\subsection{Monte Carlo simulations}
  \label{ssec:mc}

 A full Monte Carlo simulation of the \gerda\ experiment and of many of the
 related R\&D facilities has been setup in the form of a general and flexible
 framework based on {\geant}~\cite{geant03,geant06}, which is called
 \mage~\cite{mage11}. \mage\ has been widely used for \gerda-related
 simulations and background studies.  Conversely, most of the experimental
 test stands provided experimental data that were used to validate and
 benchmark {\mage}. A detailed simulation of the \textsc{LArGe} setup is also
 available within \mage.

 A few specific \gerda-related simulations were run using other codes than
 \mage. In particular, a dedicated simulation code was developed to estimate
 the residual background in the detector array due to external $\gamma$-rays,
 produced either in the surrounding rocks or in the cryostat
 volume~\cite{barabanov09}.  The simulation code SHIELD~\cite{dementyev99} was
 used to optimize the shielding required for the transportation of GeO$_{2}$
 enriched in $^{76}$Ge from the enrichment plant to the underground storage
 site~\cite{barabanov06}. Neutron spectra and fluxes produced by $\alpha$s
 from the $^{228}$Th calibration sources via the ($\alpha$,n) reactions were
 calculated through the SOURCES-4A code~\cite{codeS4A}.

%
\section{The germanium detectors}
     \label{sec:det}

 This section describes the germanium detectors that represent the core of the
 \gerda\ experiment. For Phase~I all eight detectors from the former \hdm\ and
 \igex\ experiments~\cite{hdm,igex} were refurbished and redeployed. For
 Phase~II new material amounting to 50~kg \geenr{O$_2$} and 34~kg of
 \gedep{O$_2$} was purchased. The \gedep, material depleted in \gesix\ below
 0.6~\%, was used to check the supply chain and methods of Phase~II diode
 production~\cite{depleted}. The production and characterization of the new
 detectors is ongoing.

\begin{figure}[t]
\begin{center}
\ifmakefigures%
   \includegraphics[width=0.45\columnwidth]{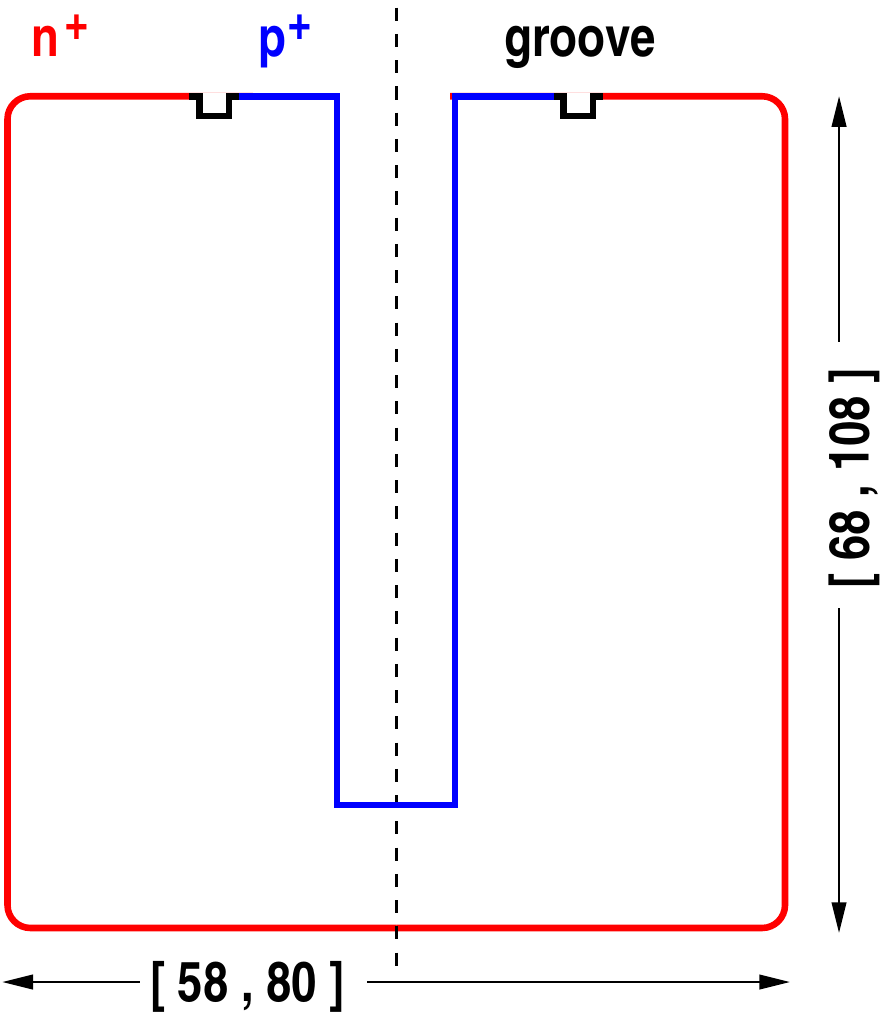}
\fi%
   \caption{	\label{fig:detschem}
      Schematic drawing of a \geenr\ diode currently operated in \gerda. The
      ranges of dimensions for the eight detectors are given in units of
      mm. The masses range from 0.98 to 2.9~kg.
  }
\end{center} 
\end{figure}

 Phase~I detectors are based on standard p-type HPGe detector technology from
 Canberra Semiconductor NV, Olen~\cite{canberra}. Standard closed-end coaxial
 detectors have a ``wrap around'' n$^+$ conductive lithium layer
 ($\sim$\,1~mm) that is separated from the boron implanted p$^+$ contact by a
 groove; the groove region is usually passivated. The detector geometry for
 one of the enriched detectors is shown schematically in
 Fig.~\ref{fig:detschem}.
 In normal DC coupled readout, the p$^+$ surface ($\sim$\,1~\mum) is connected
 to a charge sensitive amplifier and the n$^+$ surface is biased with up to
 +4600~V.  In the alternative readout scheme with AC coupling, the n$^+$
 contact is grounded and the p$^+$ contact biased with negative high voltage
 (HV).  The analog signal is still read out from the p$^+$ contact but coupled
 with a HV capacitor to the amplifier.

 Operation of bare HPGe detectors in cryogenic liquids is a non-standard
 technique. The success of \gerda\ depends strongly on the long-term stability
 of the Ge detectors operated in LAr.

\subsection{Prototype detector testing in LAr and in LN{$_2$}}
            \label{ssec:prototype}

\begin{figure*}[t]
\begin{center}
\ifmakefigures%
  \includegraphics[width=0.9\textwidth]{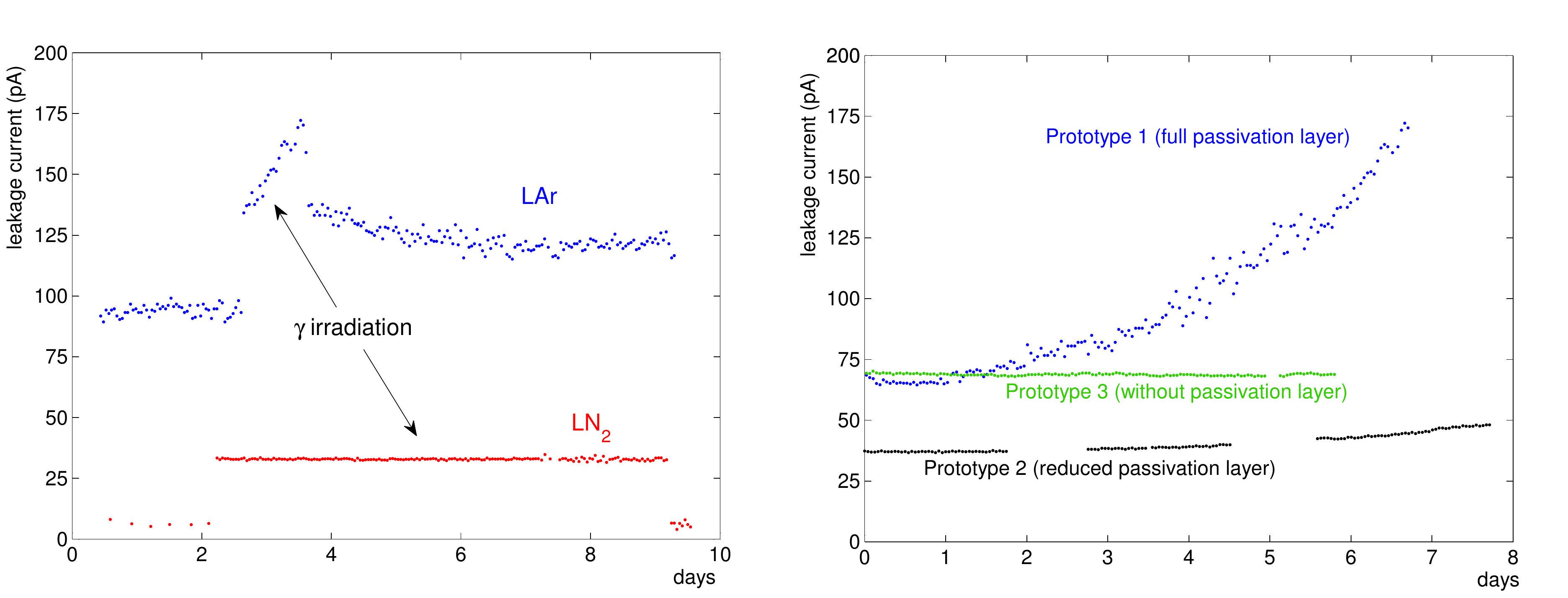}
\fi%
\caption{ \label{fig:lc}
  Left: $\gamma$-radiation induced leakage current (LC) of the first prototype
  operated in LAr.
  Right: $\gamma$-radiation induced LC for 3 prototype detectors with
  different passivation layers.
  }
\end{center}
\end{figure*}

 Before deploying the enriched detectors in LAr, bare \genat\ detectors built
 with the same technology as the Phase~I detectors were used for tests in
 GDL. A long-term study of the leakage current (LC) of bare detectors operated
 in LN$_2$ and LAr under varying $\gamma$-irradiation conditions was
 performed.  Irradiation of a first prototype detector in LAr with $\gamma$'s
 resulted in a continuous increase of the LC (see Fig.~\ref{fig:lc}, left).

 The ionizing radiation created the expected bulk current in the detector
 ($\sim$40~pA), observed as a step at the start of the $\gamma$-irradiation at
 $t\sim$2~d. This was then followed by a continuous increase of the LC. After
 about one day of irradiation, at $t\sim$3.5~d, the source was removed and the
 LC stabilized at a higher value than prior to the irradiation ($\Delta$LC
 $\approx$30~pA). No increase of the LC was observed with the same detector
 assembly in LN$_2$ after one week of irradiation.

 The process is reversible as the LC was partly restored by irradiation with
 the same source but without applying HV; the LC was completely restored to
 its initial value by warming up the detector in methanol baths. These
 measurements are the first observation of $\gamma$-radiation induced leakage
 current increase for detectors of this design operated in this way. The
 $\gamma$-radiation induced LC was measured for different HV bias values,
 source-detector configurations and HV
 polarities~\cite{marikieee,marik}. Measurements with three prototype
 detectors using different sizes of groove passivation (large area, reduced
 and none) were performed.  It was found that reducing the size of the
 passivation layer strongly suppresses the $\gamma$-radiation induced LC (see
 Fig.~\ref{fig:lc}, right).  The most likely explanation is that the LC
 increase is induced by the collection and trapping of charges produced by the
 ionization of LAr on the passivated surface of the detector.  No
 $\gamma$-radiation induced LC increase was observed with the prototype
 without passivation layer.

 For all stability measurements~\cite{marik}, the detectors were biased above
 their nominal operation voltage. The LC, continuously monitored with high
 accuracy, was at a few tens of pA for each detector, similar to the values
 measured at the detector manufacturer.  Detectors with no passivation layer
 showed the best performance in LAr.  Consequently, all \gerda\ Phase~I
 detectors were reprocessed without the evaporation of a passivation layer.
 Our positive results on the long-term stability of Ge detectors in LAr and
 LN$_2$ contradict the statements made in Ref.~\cite{kklargon}.

\subsection{Phase~I detectors}
            \label{ssec:Phasei}

 The enriched Phase~I detectors ANG~1-5 from the \hdm\ and RG~1-3 from the
 \igex\ collaborations were originally produced by ORTEC.  In addition, six
 detectors made of \genat\ are available from the GENIUS-TF
 experiment~\cite{geniustf}. They have been stored underground and therefore
 their intrinsic activity is low. Thus, they have been used in the
 commissioning phase of \gerda.  Details of the characterization of the
 enriched detectors before they were dismounted from vacuum cryostats in 2006
 are reported in Ref.~\cite{oleg}.

 The Phase~I detectors, \geenr\ and \genat, were modified at Canberra,
 Olen~\cite{canberra}, in the period from 2006 to 2008. The detector ANG~1 had
 a previous reprocessing at the same manufacturer in 1991.  The work was
 performed according to the standard manufacturer technology, however the
 passivation layer on the groove was omitted. Leakage current and capacitance
 of each detector were measured in LN$_2$ at the manufacturer site after the
 reprocessing~\cite{marik}.

\begin{table*}[hbt]
\begin{center}
\caption{ \label{tab:det}
    Characteristics of the Phase~I enriched and natural detectors. The
    isotopic abundances for \gesix, $f_{76}$, of the ANG-type detectors are
    taken from Ref.~\cite{hdmabundances}; those for RG-type detectors are from
    Ref.~\cite{igexabundances}; the natural abundance~\cite{nist} is taken for
    GTF detectors. The numbers in parentheses in the last column give the
    1$\sigma$-uncertainties (for details see Table~\ref{tab:isotopic}).
}
\vspace*{2mm}
\begin{tabular} { c | c c c c c  r}
detector &  {serial nr.} & diam. & length &
                                                total & operat.
   &  abundance \\ 
name & ORTEC      & (mm) & (mm) &mass (g)&bias (V) &   $f_{76}$~~~~~\\ \hline
ANG 1            & $^\star$)& 58.5 & ~68  & ~958 & 3200& 0.859 (13) \\
ANG 2            & P40239A & 80~~ & 107  & 2833 & 3500& 0.866 (25) \\
ANG 3            & P40270A & 78~~ & ~93  & 2391 & 3200& 0.883 (26) \\ 
ANG 4            & P40368A & 75~~ & 100  & 2372 & 3200& 0.863 (13) \\
ANG 5            & P40496A & 78.5 & 105  & 2746 & 1800& 0.856 (13) \\\hline
RG 1 ~$^\dagger$) & 28005-S & 77.5 & ~84  & 2110 & 4600& 0.8551 (10)\\
RG 2 ~$^\dagger$) & 28006-S & 77.5 & ~84  & 2166 & 4500& 0.8551 (10)\\
RG 3 ~$^\dagger$) & 28007-S & 79~~ & ~81  & 2087 & 3300& 0.8551 (10)\\\hline
GTF ~32          & P41032A & 89~~ & ~71  & 2321 & 3500& 0.078 (1)\\
GTF ~42          & P41042A & 85~~ & 82.5 & 2467 & 3000& 0.078 (1)\\ 
GTF ~44          & P41044A & 84~~ & ~84  & 2465 & 3500& 0.078 (1)\\
GTF ~45          & P41045A & 87~~ & ~75  & 2312 & 4000& 0.078 (1)\\
GTF 110          & P41110A & 84~~ & 105  & 3046 & 3000& 0.078 (1)\\
GTF 112          & P41112A & 85~~ & 100  & 2965 & 3000& 0.078 (1)\\ 
  \end{tabular} 
        \end{center}
$^\star$) produced by Canberra, serial nr. b~89002.\\
$^\dagger$) as different types of measurements vary, an uncertainty of 2~\%
             is taken in evaluations.
\end{table*}
 The detector dimensions after the reprocessing, the operating bias determined
 in the LAr test bench of GDL and with the abundance of \gesix\ measured
 earlier are reported in Table~\ref{tab:det}. A total of $\sim$300~g was
 removed from the detectors during reprocessing resulting in 17.7~kg enriched
 diodes for Phase~I. The active masses of the detectors were assessed at
 typically $\sim$87~\% by comparing $\gamma$-ray detection efficiencies to
 Monte Carlo simulations of the diodes with dead layer thicknesses
 varied~\cite{marik}.  This assessment will be refined with in-situ
 \gerda\ data.

 Cosmogenically produced isotopes $^{68}$Ge and $^{60}$Co can lead to an
 internal contamination that represents a background in the region of
 interest. The detectors are always stored at an underground facility to avoid
 exposure to cosmic rays. This applies also for the reprocessing steps, where
 the detectors were stored underground at the HADES facility~\cite{hades},
 located at a depth of about 500\,m\,w.e. at a distance of 15~km from the
 detector manufacturer. The total exposure above ground was minimized to
 $\sim$5\,days~\cite{marik}. At the start of Phase~I in November 2011, the
 estimated BI contribution from the cosmogenically produced $^{60}$Co is on
 average about $(1-2)\cdot10^{-3}$~\ctsper. The bulk of the $^{60}$Co activity
 comes from the production before the underground installation of the
 detectors for the \hdm\ and \igex\ experiments. The contribution from
 $^{68}$Ge is negligible since it decayed away.

\begin{figure}[H!b]
\begin{center}
\ifmakefigures%
   \includegraphics[width=\columnwidth]{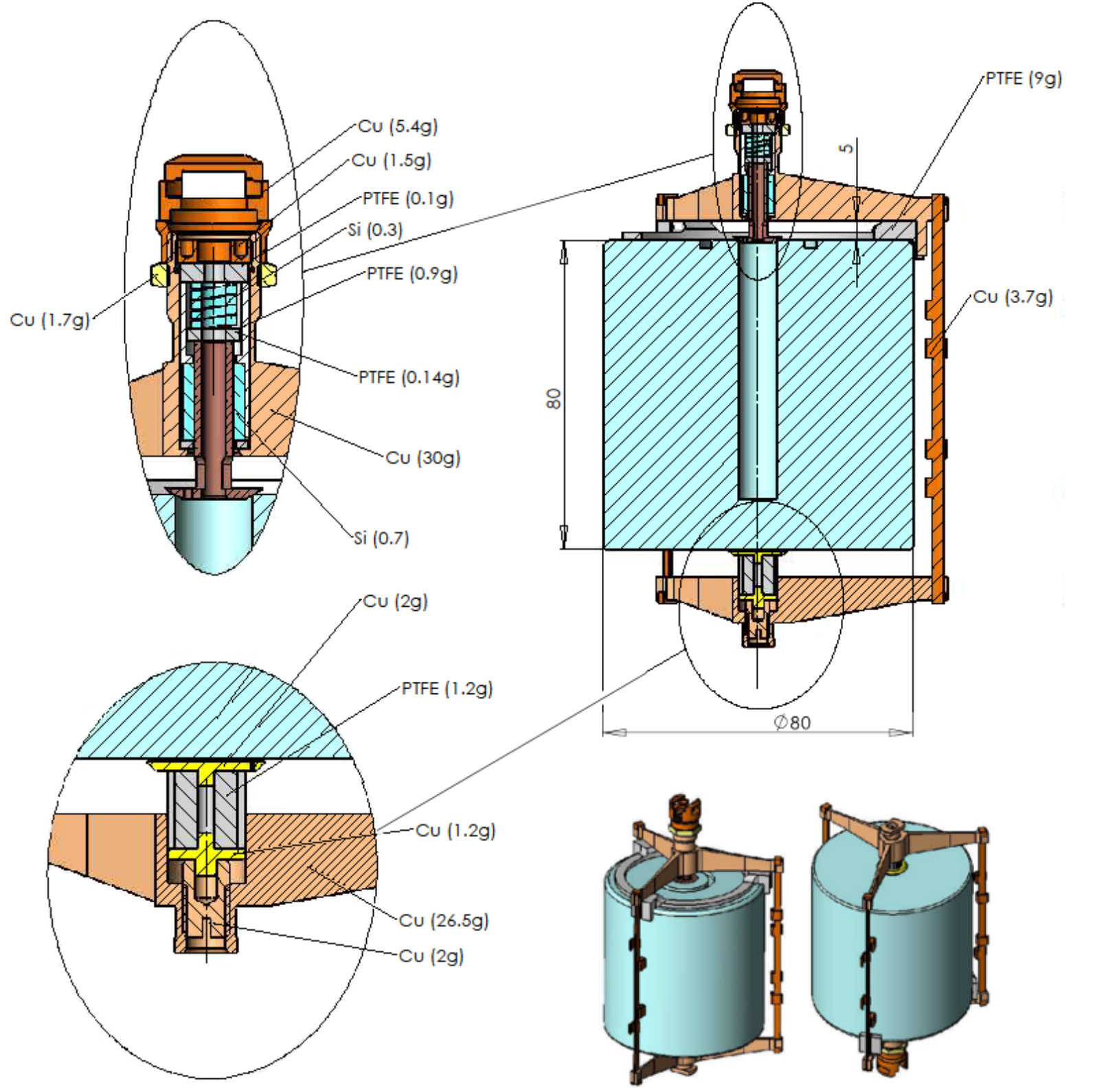}
\fi%
   \caption{ \label{fig:detsupp}
    Drawing of a Phase~I detector assembly. The signal contact is realized by
    a conical copper piece (``Chinese hat'') that is pushed by a silicon spring
    onto the p$^+$ contact (inset left top). High voltage is applied to the
    n$^+$ contact by a copper strip (not shown) pressed by a copper disc which
    in turn is electrically insulated by a PTFE cylinder (inset bottom
    left). The force to achieve good electrical contact is actuated through a
    copper screw.  Masses and dimensions of the assembly are given for the RG3
    detector.
  }
\end{center} 
\end{figure}

 The mounting scheme of the detectors has competing requirements. It must have
 a low mass to minimize sources of radiation near to the detectors. However,
 the construction must be sufficiently sturdy to provide safe suspension. It
 must support the cables for detector bias and readout. Furthermore, the
 diodes must remain electrically isolated from all other materials. The chosen
 support design is depicted in Fig.~\ref{fig:detsupp} where the contacting
 scheme is shown as well. In order to reach the background goals of \gerda,
 the amount of material is minimized.
 Only selected high radiopurity materials were used: copper
 ($\sim$80~g), PTFE ($\sim$10~g), and silicon ($\sim$1~g). The results of the
 $\gamma$ ray spectroscopy measurements (see sec.~\ref{ssec:screening}),
 combined with Monte Carlo simulations give an upper limit on the BI
 contribution from the detector support of $\leq\,10^{-3}$~\ctsper.

 One of the prototype detectors was mounted in a support of the Phase~I design
 to test the electrical and mechanical performance. This confirmed the
 mounting procedure, the mechanical stability, the signal and HV contact
 quality, and the spectroscopic performance of this design. During this test,
 the energy resolution was the same as was achieved previously when the same
 detector was mounted in a standard vacuum cryostat, i.e. $\sim$2.2~keV full
 width at half maximum (FWHM) at the 1332~keV spectral line of $^{60}$Co.

\begin{figure}[bh]
\begin{center}
\ifmakefigures%
   \includegraphics[width=\columnwidth]{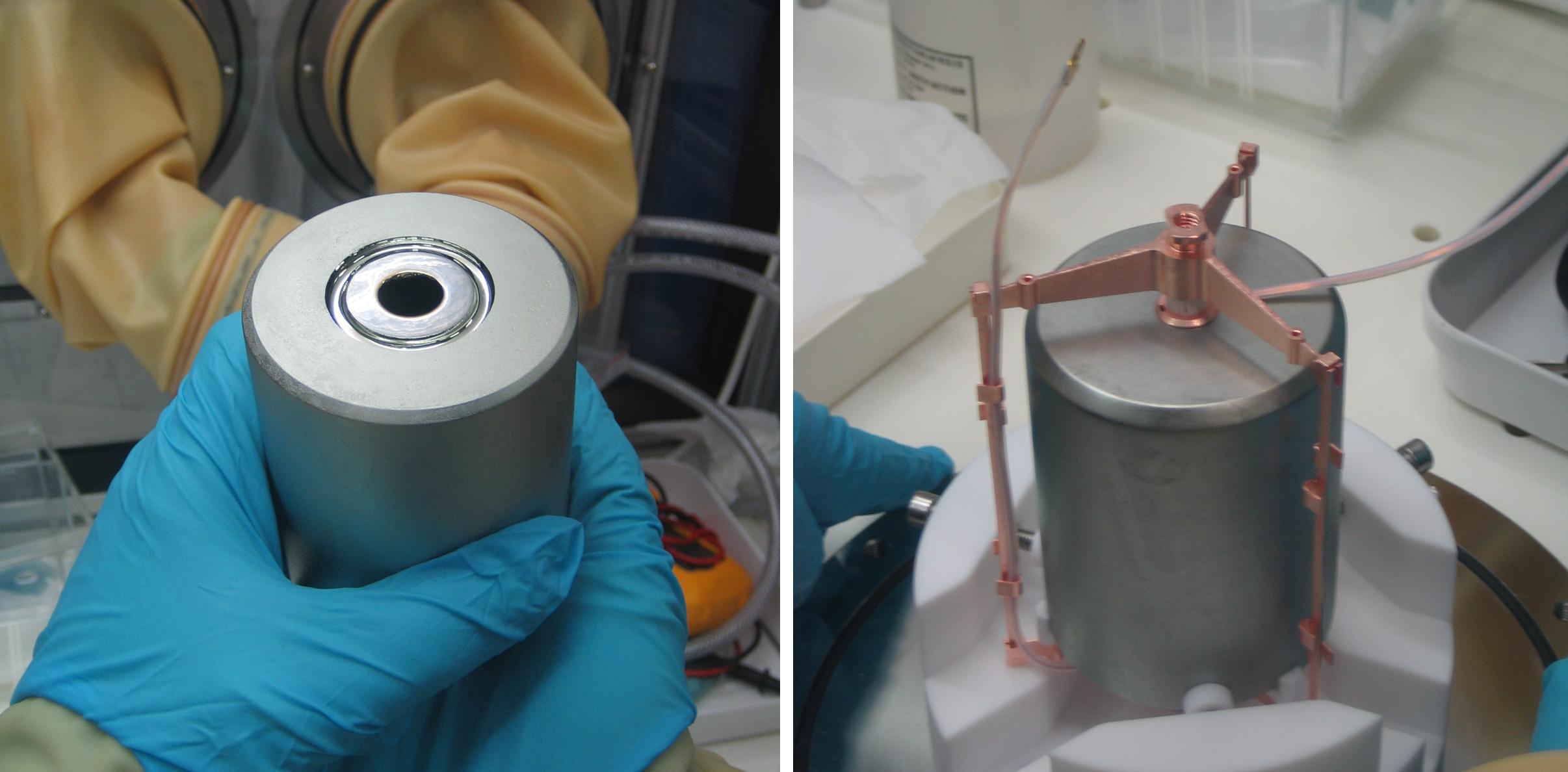}
\fi%
   \caption{	\label{fig:detcan}
   Left: A Phase~I detector after reprocessing at Canberra, Olen. The
         conductive lithium layer (n$^+$ contact) and the boron implanted bore
         hole (p$^+$ contact) are separated by a groove.
   Right: The detector is mounted upside down in a low-mass holder (groove no
         longer visible).
  }
\end{center} 
\end{figure}

 Fig.~\ref{fig:detcan} shows one of the Phase~I detectors before and after
 mounting in its custom made support structure.  The Phase~I detectors were
 mounted in their final low-mass supports in 2008 and their performance
 parameters (leakage current, counting efficiency, energy resolution) were
 measured in LAr as a function of bias voltage~\cite{marik}. The detector
 handling was performed in GDL entirely within an environment of N$_2$ gas.
 The LC of the majority of the detectors was at the same level as measured at
 the detector manufacturer after reprocessing. The detectors ANG~1, ANG~3 and
 RG~3 showed high LCs even after successive thermal cycling and required
 additional reprocessing to reach an acceptable performance. Spectroscopic
 measurements were performed, as described in Ref.~\cite{jinst}, with the
 preamplifier mounted in a gaseous Ar environment in the neck of the LAr
 cryostat. The energy resolutions of the Phase~I detectors was between 2.5 and
 5.1~keV (FWHM) for the 1332~keV spectral line of $^{60}$Co. An improvement of
 the energy resolution of the detectors was observed after polishing
 the diode surface in the location of the HV contact.

 Since November 2011 all the enriched Phase~I detectors have been inserted
 into the \gerda\ cryostat.

\subsection{Phase~II detectors}
            \label{ssec:Phaseii}

\begin{table*}[th]
\begin{center}
\caption{\label{tab:isotopic}
     The relative number of nuclei for the different isotopes is shown for the
     different detector batches.  The isotopic 
     composition of the depleted material is the average of measurements by
     the collaboration and ECP; that for natural germanium is given for
     comparison.
}
\vspace*{2mm}
\begin{tabular}{cr|lllll}
&&\multicolumn{5}{c}{germanium isotope}\\ \cline{3-7}
detector batch &Ref.&\multicolumn{1}{c}{70} &\multicolumn{1}{c}{72} 
 &\multicolumn{1}{c}{73} &\multicolumn{1}{c}{74} &\multicolumn{1}{c}{76} \\
\hline
natural &\cite{nist}
                 & 0.204(2) & 0.273(3)  & 0.078(1)  & 0.367(2)    & 0.078(1) \\
\hdm\ -- ANG~1 &\cite{zuber}
                 & 0.0031(2)& 0.0046(19)& 0.0025(8) & 0.131(24)   & 0.859(29)\\
\igex &\cite{igexabundances}
                 & 0.0044(1)& 0.0060(1) & 0.0016(1) & 0.1329(1)  & 0.8551(10)\\
\gerda\ depleted &
                 & 0.225(2) & 0.301(3)  & 0.083(1)  & 0.390(5)    & 0.006(2) \\
\gerda\ Phase~II $^\star$)& \cite{ecp} 
                 & 0.0002(1)& 0.0007(3) & 0.0016(2) &0.124(4)   & 0.874(5)\\
\majorana &\cite{frankpriv}
                 & 0.00006  & 0.00011   & 0.0003    & 0.0865     & 0.914\\
\end{tabular}
\end{center}
$^\star$) numbers in brackets represent the range of measurements from ECP.
\end{table*}

 In order to increase the active mass a new set of enriched germanium
 detectors is currently in production for Phase~II of \gerda.  A brief
 description of the activities is given here.

 A batch of $37.5$~kg of \geenr\ was procured by the Electrochemical Plant
 (ECP) in Zhelenogorsk, Russia~\cite{ecp} in 2005. The isotopic content of the
 enriched germanium is given in Table~\ref{tab:isotopic}.  The enrichment was
 performed by centrifugal separation of GeF$_4$ gas, and the \geenr\ was
 delivered in the form of 50~kg \geenr{O$_2$}.

 A major concern during all steps is the production of long-lived
 radioisotopes via cosmogenic activation, in particular $^{68}$Ge and
 $^{60}$Co.  Specially designed containers were used to transport the
 material~\cite{barabanov06} by truck from Siberia to Germany; the
 {\geenr}O$_2$ was then kept in the HADES facility in underground storage
 while not being processed.

 A series of reduction and purification tests with \gedep\ was organized. A
 complete test of the production chain from enrichment to the tests of working
 diodes was performed within a year.  Based on results on isotopic dilution
 and yield, it was decided to further process the material at PPM Pure metal
 GmbH~\cite{ppm}.  The processing of the $^{\rm enr}$GeO$_2$ took place in
 spring 2010. The steps included a reduction of GeO$_2$ to ``metallic'' Ge,
 with typical purity of 3N ($99.9$~\% Ge) and then zone refinement to 6N
 purity, corresponding to $\geq 99.9999$~\% chemical purity in Ge.  After
 reduction $37.2$~kg of germanium metal remained. From this material,
 $36.7$~kg of germanium remained after zone refinement, $35.5$~kg of which
 satisfies the 6N requirement. The biggest loss of material came from the
 etching of the reduced metal.  The material was stored in a mining museum
 near PPM between processing steps.

 For further processing the material was shipped in a special container to
 Canberra, Oak Ridge~\cite{oakridge}.  Zone refining to 11N and pulling
 crystals of the required dimensions with a net carrier concentration
 corresponding to 12N purity and other characteristics such as crystal
 dislocation density within a specified range~\cite{ref:Haller} has been
 completed there.  The crystals have been cut and 30 slices have been brought
 to Canberra, Olen, for detector production.  The total mass of the slices
 amounts to 20.8~kg.

 The new detectors are of Broad Energy Germanium (BEGe)~\cite{begeCanberra}
 type with good pulse shape discrimination properties~\cite{dusan09,matteo11}.
 The first seven have been produced and tested in vacuum cryostats
 reaching an energy resolution of 1.7~keV FWHM at the 1332~keV $^{60}$Co line.
 Tests in LN$_2$ and LAr are underway.  Five of them have been placed into a
 string and inserted into the \gerda\ cryostat in July 2011.

\section{Experimental setup }
 \label{sec:setup}

 \gerda\ occupies an area of 10.5$\times$10.4 m$^2$ in Hall~A of
 \LNGS\ between the TIR tunnel and the LVD experiment. A model of the
 experiment is shown in Fig.~\ref{fig:overview}. The floor area has been
 refurbished with reinforced concrete for enhanced integral stability and was
 sealed with epoxy for water tightness. A grid surrounding the water tank is
 connected to the new \LNGS\ water collection system.  The various components
 were erected sequentially. The construction of the bottom plate of the water
 tank (sec.~\ref{ssec:water}) was followed by the installation of the cryostat
 (sec.~\ref{ssec:cryo}) which arrived by a flat-bed truck from the
 manufacturer in March 2008. After the acceptance tests, the water tank
 construction was resumed and finished in June 2008. Subsequently the
 \gerda\ building (sec.~\ref{ssec:superstructure}) was built and on top of it
 the clean room (sec.~\ref{ssec:clean}) was erected; the latter houses the
 lock system with a glove box, the calibration system (sec.~\ref{ssec:calib})
 as well as auxiliary cabinets.  The earthquake tolerance of the setup was
 verified by calculating the relative motions of cryostat, water tank and
 \gerda\ building for a seismic event with strength and frequency parameters
 provided by \LNGS.  The muon veto system (sec.~\ref{ssec:muonveto}) consists
 of two parts, the water Cherenkov detector which is mounted within the water
 tank and an array of plastic scintillators which are located on the roof of
 the clean room.

\subsection{The cryostat and its cryogenic system}
\label{ssec:cryo}

 The \gerda\ cryostat holds 64~m$^3$ of LAr which serves as medium for the
 cryogenic operation of the bare Ge diodes as well as a shield against the
 remnants of the external $\gamma$ background penetrating the surrounding
 water and against the radioactivity of the cryostat itself.  Leakage of radon
 from the atmosphere into the cryostat is prevented by the exclusive use of
 metal seals in the joints and valves and by keeping an overpressure of about
 \cpowten{3}{4}~Pa against atmosphere.  In the original design copper of low
 radioactivity, i.e.~$<$\,20~\mubq/kg of $^{228}$Th, was foreseen as
 production material. However, safety issues and an unexpected cost
 increase forced the change to a stainless steel cryostat with an internal
 copper shield. Taking into account the measured radioactivity values of the
 stainless steel material~\cite{steel_paper} (see sec.~\ref{ssec:screening}),
 the thickness of the copper shield was determined by analytical calculations
 and MC simulations such that sources of $\gamma$ radiation external to the
 cryostat and the cryostat itself contribute to the BI by about
 \cpowten{0.5}{-4}~\ctsper~\cite{barabanov09}.

 This section describes the cryostat and the cryogenic system required for its
 stable operation and some performance features of the setup. At the end
 special safety aspects are discussed that result from the operation of a
 cryostat immersed into a large water volume located in an underground site.

\subsubsection{The cryostat}
 \label{sssec:cryostat}

\begin{figure}[bth!]
\begin{center}
\ifmakefigures%
  \includegraphics[width=\columnwidth]{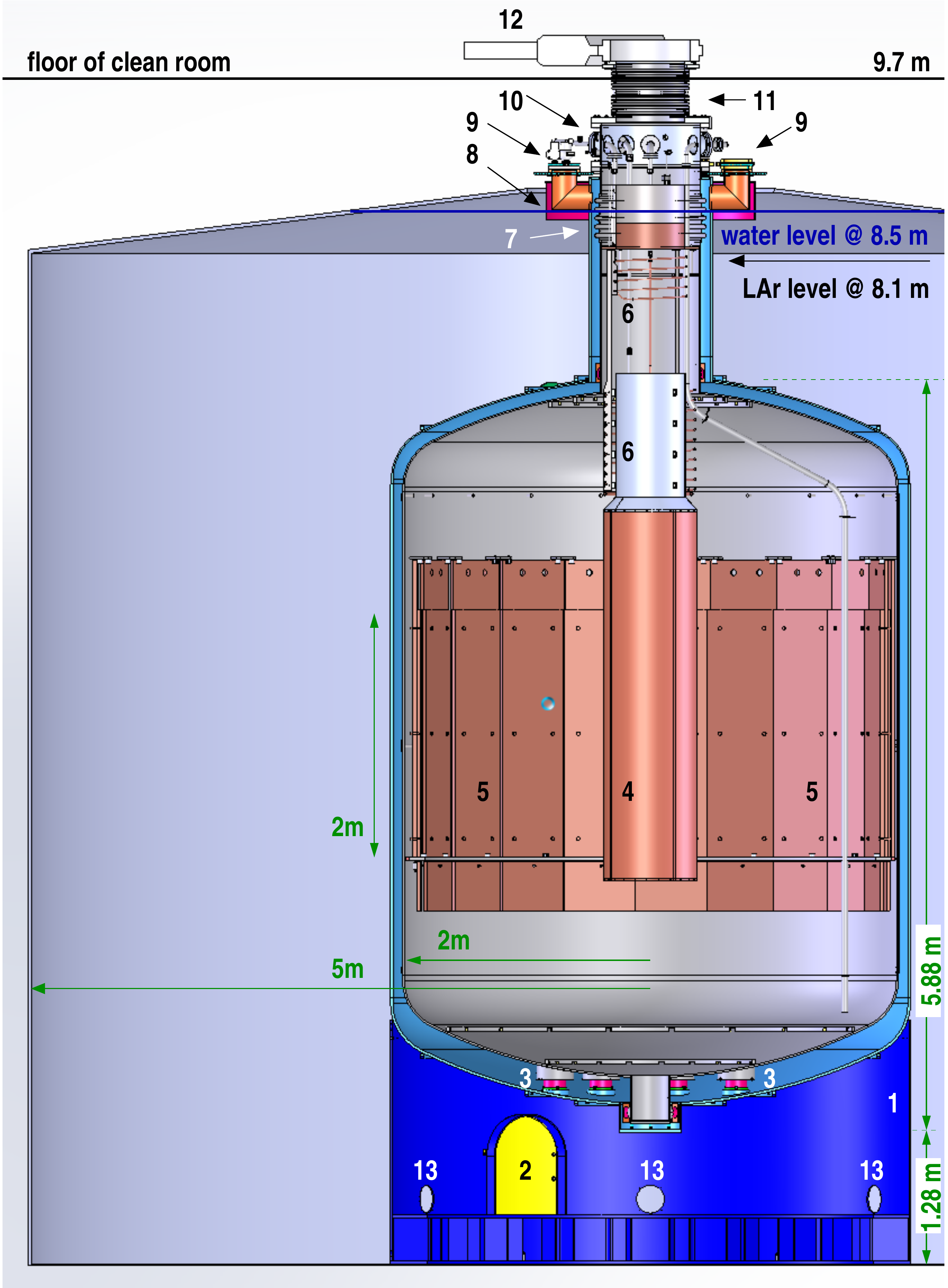}
\fi%
\caption{\label{fig:cryostat}
  Cross section of the  LAr cryostat inside the water tank (right part
  cut away).  
  The following components are indicated: skirt (1), access hole (2), Torlon
  support pads (3), radon shroud (4), internal copper shield (5), lower and
  upper heat exchanger (6), bellow in neck of inner vessel (7), balcony (8),
  DN200 ports (9), manifold (10), bellow between cryostat and lock (11) and
  DN630 shutter (12). The skirt provides 6 mounts for PMTs (13).
  }
\end{center}
\end{figure}

 The cross section of the super-insulated cryostat is shown in
 Fig.~\ref{fig:cryostat}.
 The cryostat is supported by a skirt (item~1) at a height of about 1.3~m
 above the bottom of the water tank.  Access to the volume below the cryostat
 within the skirt is provided by two manholes (item~2).  The cryostat consists
 of two coaxial vessels comprising of torospherical heads of 4200 and 4000~mm
 outer diameter and corresponding cylindrical shells of about 4~m height. The
 inner vessel rests on eight Torlon~\cite{torlon} pads (item~3) located on the
 bottom head of the outer vessel.  Both vessels have a cylindrical neck of
 1.7~m height and are connected at the top.  The compensation for thermal
 shrinkage of the inner container is provided by a double-walled stainless
 steel bellow in its neck (item~7). In the upper region the outer neck carries
 four DN200 flanges (item~9) which are protected against the water by a kind
 of ``balcony'' surrounding the neck (item~8).  A flexible rubber fabric
 closes the gap between the water tank roof and the balcony.  The flanges
 allow access to the volume between inner and outer vessel and they carry the
 pump and the pressure sensors for the insulation vacuum as well as a safety
 disc as protection against overpressure.  The neck of the inner vessel with
 an inner diameter of 800~mm provides the only access to the interior of the
 cryostat. A manifold (item~10) on top of the neck carries the flanges for the
 feedthroughs of all devices that penetrate into the cold volume including a
 filling tube, gas exhaust tube, tubes for active cooling, and feedthroughs
 for the cryostat instrumentation. The Ge diodes are lowered into the cryostat
 through a lock which resides in the clean room above the manifold (see
 sec.~\ref{ssec:clean}). Relative movements between manifold and lock are
 decoupled from each other with a flexible bellow of 600~mm diameter
 (item~11). A DN630 UHV shutter (item~12) on top of the bellow allows the
 stand-alone operation of the cryostat without lock.

 The internal copper shield (item~5) consists of sixty 3~cm thick overlapping
 plates of high purity oxygen free radiopure (OFRP) copper with a total mass of
 16~t. They are mounted on a support ring achieving a copper thickness of 6~cm
 for the central 2~m high ring (centered at 4~m height) and of 3~cm thickness
 in a range of 40~cm above and below.

 Radon can emanate from the vessel walls and may be transported by convection
 close to the Ge diodes.  To prevent this a central volume of about 3~m height
 and 750~mm diameter is separated from the rest by a cylinder (item~4) made
 out of 30~\mum\ thick copper foil. This cylinder is called the radon shroud.

 During production and after its deployment at \LNGS\ the cryostat has been
 subjected to several acceptance and performance tests. Both the inner and the
 outer vessel passed the pressure vessel tests according to the European
 pressure vessel code PED 97/23/EC. Helium leak tests for the inner and the
 outer vessel showed no leak at the 10$^{-5}$~(Pa$\cdot\ell$)/s
 range. Evaporation tests with LN$_2$ established the specified thermal loss
 of $<$\,300~W both at the factory and after delivery. The $^{222}$Rn
 emanation rate of the inner volume of the cryostat has been measured at room
 temperature at several stages with the MoREx system~\cite{morex} (for details
 see Table~\ref{tab:rn_cryostat} in sec.~\ref{sssec:rn_cryo}). After iterated
 cleaning the empty cryostat exhibited the excellent value of $(14\pm4)$~mBq
 which increased after the mounting of the Cu shield and the cryogenic
 instrumentation by about 20~mBq at each step, leading to a final value of
 $(54.7\pm3.5)$~mBq. A uniform distribution of this amount of $^{222}$Rn in
 the LAr would correspond to a BI$\sim$7$\cdot$10$^{-4}$~\ctsper.  Depending
 on its tightness, the radon shroud is expected to reduce this contribution by
 up to a factor of seven.

\subsubsection{Cryogenic system}
    \label{sssec:cryogenics}

 The cryogenic infrastructure consists of storage tanks, super-insulated
 piping, and the systems for vacuum insulation, active cooling, process
 control, and exhaust gas heating. The power for the entire system is taken
 from a dedicated line which is backed-up by the \LNGS\ diesel rotary
 uninterruptible power supply.

 The storage tanks for LN$_2$ and LAr, about 6~m$^3$ each, are located at
 about 30~m distance. To minimize argon losses they are connected by a
 triaxial super-insulated pipe (LAr, LN$_2$ and vacuum super-insulation from
 inside to outside) to the cryostat. The LAr tank has been selected for low
 radon emanation. The tank has been used for the filling of the cryostat and
 will be used further for optional refillings. The LAr passes through a
 LN$_2$-cooled filter filled with synthetic charcoal~\cite{charcoal} to retain
 radon as well as through two PTFE filters with 50~nm pore size to retain
 particles. For the first filling the charcoal filter was bypassed.

 The insulation vacuum has to be maintained in a volume of about
 8~m$^3$. Out-gassing materials in this volume include about 75~m$^2$ of
 multilayer insulation and 50~m$^2$ of additional thermal insulation
 (Makrolon~\cite{makrolon} of 6~mm thickness). A pressure of 10$^{-3}$~Pa was
 reached after two months of pumping with a turbo pump of 550~$\ell$/s pumping
 speed and intermediate purging with dry nitrogen. After cool down the
 pressure dropped to about 2$\cdot$10$^{-6}$~Pa.  At a residual out-gassing
 rate in the range of 10$^{-5}$~(Pa$\cdot\ell$)/s, the turbo pump is kept
 running continuously.

 The active cooling system uses LN$_2$ as cooling medium. It has been
 designed~\cite{Hstroh} to subcool the main LAr volume in order to minimize
 microphonic noise in the cryostat while maintaining a constant (adjustable)
 working pressure without evaporation losses.  This is accomplished by two
 LN$_2$/LAr heat exchangers (item~6 in Fig.~\ref{fig:cryostat}), spirals of
 copper tube located in the main volume and at the liquid/gas surface in the
 neck, respectively. With the nitrogen gas pressure of \cpowten{1.2}{5}~Pa
 absolute, corresponding to a LN$_2$ boiling temperature of 79.6~K, the LAr is
 cooled to about 88.8~K. Since the temperature is slightly higher than the
 boiling point at standard atmospheric pressure, the cryostat builds a slight
 overpressure until an equilibrium is reached such that no argon is lost. The
 daily LN$_2$ consumption is about 280~$\ell$.

 In case of an incident like the loss of insulation vacuum, LAr will evaporate
 at an estimated rate of up to 4.5~kg/s. The cold gas has to be heated to a
 temperature above 0$^\circ$C before it is discharged to the
 \LNGS\ ventilation system. This is achieved by a water-gas heat exchanger
 (see Fig.~\ref{fig:wt-plant}) using the \LNGS\ cooling water or the
 \gerda\ water tank reservoir.

 Complete control over almost all processes is achieved with a programmable
 logic controller (PLC) Simatic~S7 from Siemens which continuously monitors
 the information provided by more than 10 redundant pairs of Pt100 temperature
 sensors distributed in the cryostat volume, the vacuum gauges, and the level
 and pressure sensors. To improve the safety further pressure regulation was
 installed, that is independent of the PLC.  The output of a stand-alone
 pressure gauge (SMAR LD301, \cite{smar}) regulates directly the positioner of
 a valve.  Two such systems are implemented to further increase the
 reliability.  All status information is communicated to the general
 \gerda\ slow control system (sec.~\ref{ssec:slow}) and can be accessed
 globally via a web-based graphical user interface that also allows restricted
 remote control.
 
 Since its filling with LAr in December 2009, the cryostat has remained at LAr
 temperature and operations have been stable. Except for a small refill of LAr
 during the tuning of the active cooling system and one more following a
 forced Ar evaporation for a radon measurement in the exhaust gas, no
 additional LAr refill was necessary.

\subsubsection{Safety considerations}
\label{sssec:cryosafety}

 The additional risks of operating a cryostat within a water tank due to the
 huge latent water heat were analyzed early in the design phase. Specific
 mitigation measures were realized in the design, construction and the
 operation of the cryostat and cryogenic system. The most important ones are
 summarized below.

 The cryostat was designed and produced according to the European pressure
 vessel code for a nominal overpressure of \cpowten{1.5}{5}~Pa, even though it
 is operated below the limit of \cpowten{0.5}{5}~Pa above which this code
 applies. An additional safety margin is provided by an increase of the wall
 thickness of the cold vessel by 3~mm. The risk for any leak in one of the
 vessel's walls is further reduced by the lack of any penetrations in the
 inner or outer vessel below the water fill level, the 100\,\% X-raying of the
 welds and an earthquake tolerance of 0.6~g.  The use of ductile construction
 materials guarantees the cryostat to follow the leak-before-break principle.
 In case of a leak, the implementation of a passive insulation at the outside
 of the inner and the outer vessel will limit the evaporation rate to a
 tolerable maximum of about 4.5~kg/s.

 The oxygen fraction in air is monitored continuously for any low level
 employing several units placed in the \gerda\ building and in the clean room.
 Further enhanced safety features include full redundancy of pressure and
 level sensors as well as the use of both a rupture disk and a safety valve
 for overpressure protection.  The insulation vacuum is continuously monitored
 with a residual gas analyzer reading the partial pressures for water, argon,
 and nitrogen. This information will be used for diagnostics in case of an
 unexpected rise in total pressure. In case of a relevant leak the PLC would
 automatically start the drainage of the water tank.  A realistic test has
 established the complete drainage to be possible within less than two hours
 (see sec.~\ref{ssec:water}).
%
\subsection{The water tank and its water plant}
 \label{ssec:water}

 The water tank when filled with water provides a 3~m thick water buffer
 around the cryostat whose purpose is fourfold: ($i$) to moderate and absorb
 neutrons, ($ii$) to attenuate the flux of external $\gamma$ radiation,
 ($iii$) to serve as Cherenkov medium for the detection of muons crossing the
 experiment, and ($iv$) to provide a back-up for the \LNGS\ cooling water
 which in case of emergency might be needed to heat the argon exhaust gas.
     
\subsubsection {The water tank}
 \label{sssec:watertank}

\begin{figure}[b]
\begin{center}
\ifmakefigures%
     \includegraphics[width=\columnwidth]{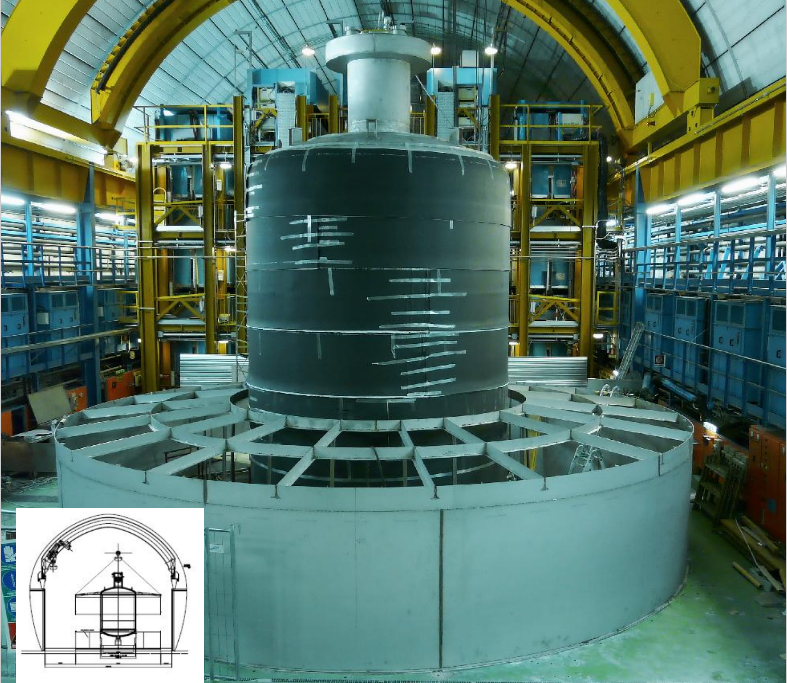}
\fi%
\caption{\label{fig:wt-constr}
    The water tank under construction in Hall A of \LNGS\ in front of the LVD
    detector.  The inset shows how the tank is assembled from top to bottom.
    The hall crane lifts the upper part to which
    another cylinder segment of about 2~m height is welded. The cryostat 
    in the center is protected by a black foamed plastic during the
    construction of the water tank.
}
\end{center}
\end{figure}

 The \WT\ with a nominal capacity of 590\,\cum\ was designed following the
 API~650 regulation and according to the Eurocode~8 for the design of
 structures for earthquake resistance.  It was built completely on site after
 the installation of the cryostat on the pre-installed butt-welded ground
 plate (Fig.~\ref{fig:wt-constr}).
 It consists of a cylinder of 10~m diameter and 8.3~m height covered by a
 conically-shaped roof which extends up to 8.9~m; the water level is kept at
 8.5~m. AISI 304L stainless steel was used exclusively as construction
 material. The sheet metal plates for the cylindrical shell have a thickness
 from 7~mm to 5~mm and are joined by butt welds using externally (internally)
 MIG (TIG) welding. An additional bottom reinforcement has been applied at the
 1 foot level.  Following the UNI~EN~1435 code, a significant fraction of the
 400~m length of welds was X-ray tested.

 Access into the \WT\ for the installation and maintenance of the muon veto
 (sec.~\ref{ssec:muonveto}) is possible through a manhole at the bottom of
 1400$\times$800~mm$^2$ size.  The roof has a central hole of 1200~mm diameter
 through which the neck of the cryostat sticks out.  The gap between neck and
 the roof is closed by a flexible membrane made of rubber to block radon and
 light from the water volume.  Radon intrusion is further reduced by a
 slightly over-pressurized nitrogen blanket between water and roof. Besides
 numerous small flanges, the \WT\ has a further DN600 manhole as well as a
 DN600 chimney for the PMT cables on the roof, and, at the bottom, two DN300
 flanges for fast water drainage.

 The water tank was filled via a dedicated pipeline from the Borexino
 plant~\cite{borexinowp} with ultrapure water of resistivity close to the
 physical limit of 0.18~M$\Omega\cdot$m.  The static test of the water tank
 consisted in the measurement of its radial deformation of the tank as
 function of the water column height and finally applying an overpressure of
 \powten{4}~Pa. Radial deformations were measured in three azimuthal locations
 at a height of 1~m and in one location at a height of 4~m. The maximum
 deformation was 7 to 8~mm as measured both in the azimuth of the manhole at
 1~m height and on the opposite side of the tank at 4~m height. The
 deformations were proven to be elastic.

\begin{figure}[h!b]
\begin{center}
\ifmakefigures%
      \includegraphics[width=\columnwidth]{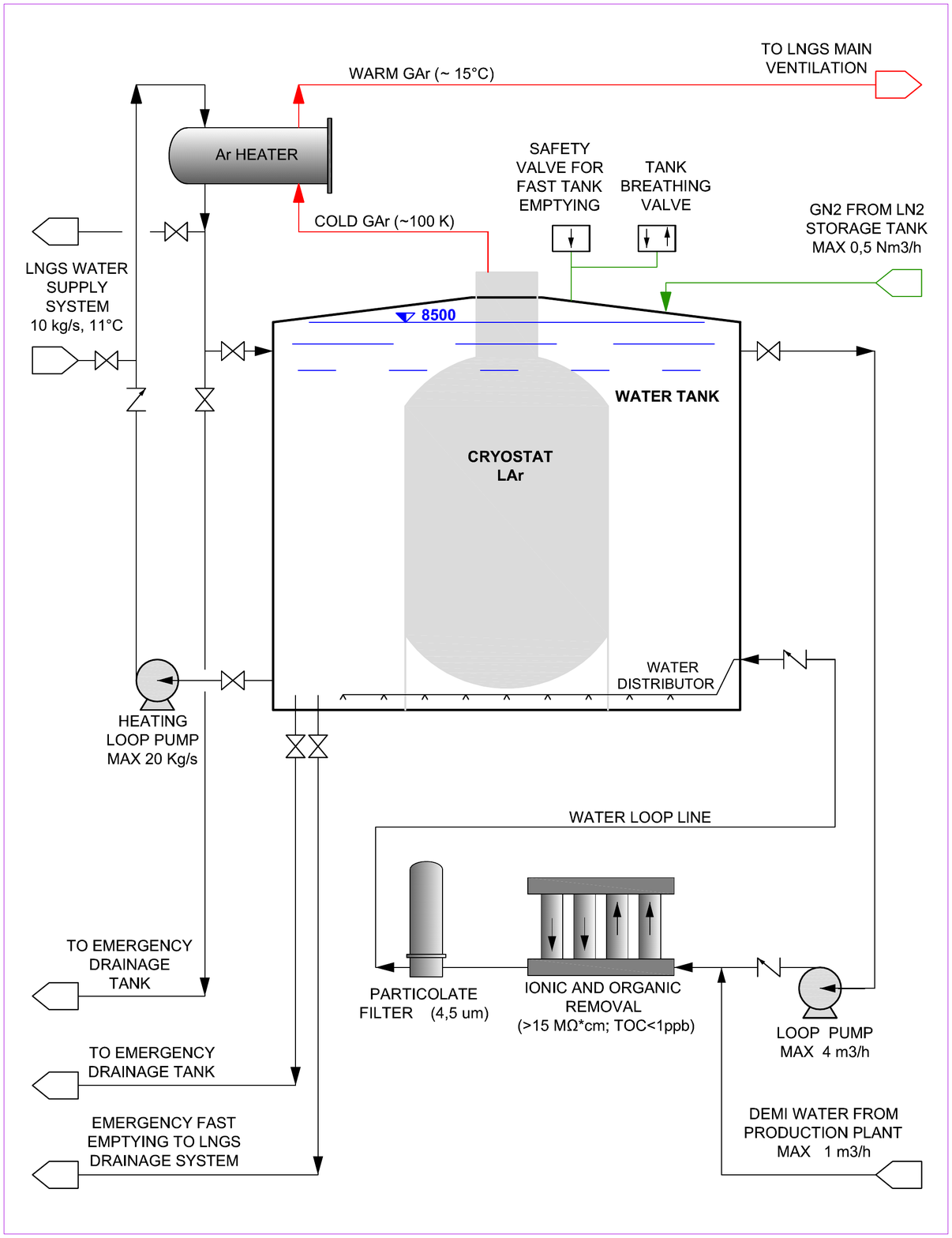}
\fi%
\caption{\label{fig:wt-plant}
     Schematic of the \gerda\ water system including the drainage,
      the argon exhaust gas heater and the  water plant.
}
\end{center}
\end{figure}

 The \WT\ exhibits various features to ensure safe operation (see
 Fig.~\ref{fig:wt-plant}).  A pressure relief valve will open when the nominal
 overpressure of \cpowten{(2-3)}{3}\,Pa is exceeded. Complete drainage of the
 water was demonstrated in less than two hours. A constant drainage rate
 through a new DN250 pipe underneath the TIR tunnel of up to 65~$\ell$/s is
 controlled by the PLC. According to the actual water level, the PLC sets the
 opening angle of a butterfly valve on that pipe to control the rate. A second
 pipe, with a maximum flow rate of 16~$\ell$/s, leads via the grid to the
 Hall~A pits that are devoted to collecting any fluid accidentally discharged
 by the experiments. In an emergency, a third channel is opened to pump water
 from the water tank at a rate of 20~$\ell$/s through the heater for the LAr
 exhaust gas (Fig.~\ref{fig:wt-plant}). This third channel also drains to the
 pits in Hall~A. During such an emergency event, an additional safety valve
 opens a vent to prevent a collapse of the water tank.

\subsubsection{The water plant}
   \label{sssec:waterplant}

 The water plant (Fig.~\ref{fig:wt-plant}) has the function to keep
 the fraction of ions normally existing in the water, especially U, Th, K, as
 low as possible (fractions of ppm). Also the level of the Total Organic
 Carbon (TOC) must be controlled, otherwise they would cause a gradual
 degradation in the optical transparency of the water over time.

 The water in the \gerda\ tank is kept in constant circulation by a loop pump
 at typically 3~m$^3$/h.  In its return path the water is purified by an
 ``Ultra-Q'' unit. This is a special device equipped with four disposable
 cartridges containing specific resins, that removes TOC and ions (both
 anionic and cationic) from the water.  Finally, the water is filtered for the
 removal of suspended particles and returned to the bottom of the \WT\ via a
 circular distribution system.  The quality of the water is monitored after
 the filter by its resistivity and is typically higher than
 0.17~M$\Omega\cdot$m.  The high light yield observed in the muon veto system
 (sec.~\ref{ssec:muonveto}) is further proof of the excellent water quality.

\subsection {The \gerda\ building}
   \label{ssec:superstructure}

 The \gerda\ building evolved from the need of a superstructure that supports
 a platform above the water tank and cryostat to host a clean room with the
 lock system for the insertion of the Ge diodes into the cryostat.  The blue
 beams of the superstructure are visible surrounding the water tank in
 Fig.~\ref{fig:overview}.  The gap between the \WT\ and \LVD\ is occupied by
 laboratory rooms on three levels plus a platform and a staircase.  The ground
 floor houses the water plant and a radon monitor, the first floor two control
 rooms (one of them dedicated to \LVD) and the second floor part of the
 cryogenic infrastructure including the heater for the Ar exhaust gas, safety
 valves and PLC as well as the electronics for the muon veto.
\subsection{The clean room, twin lock and detector suspension systems}
 \label{ssec:clean}

\begin{figure}[h]
\begin{center}
\ifmakefigures%
      \includegraphics[width=\columnwidth]{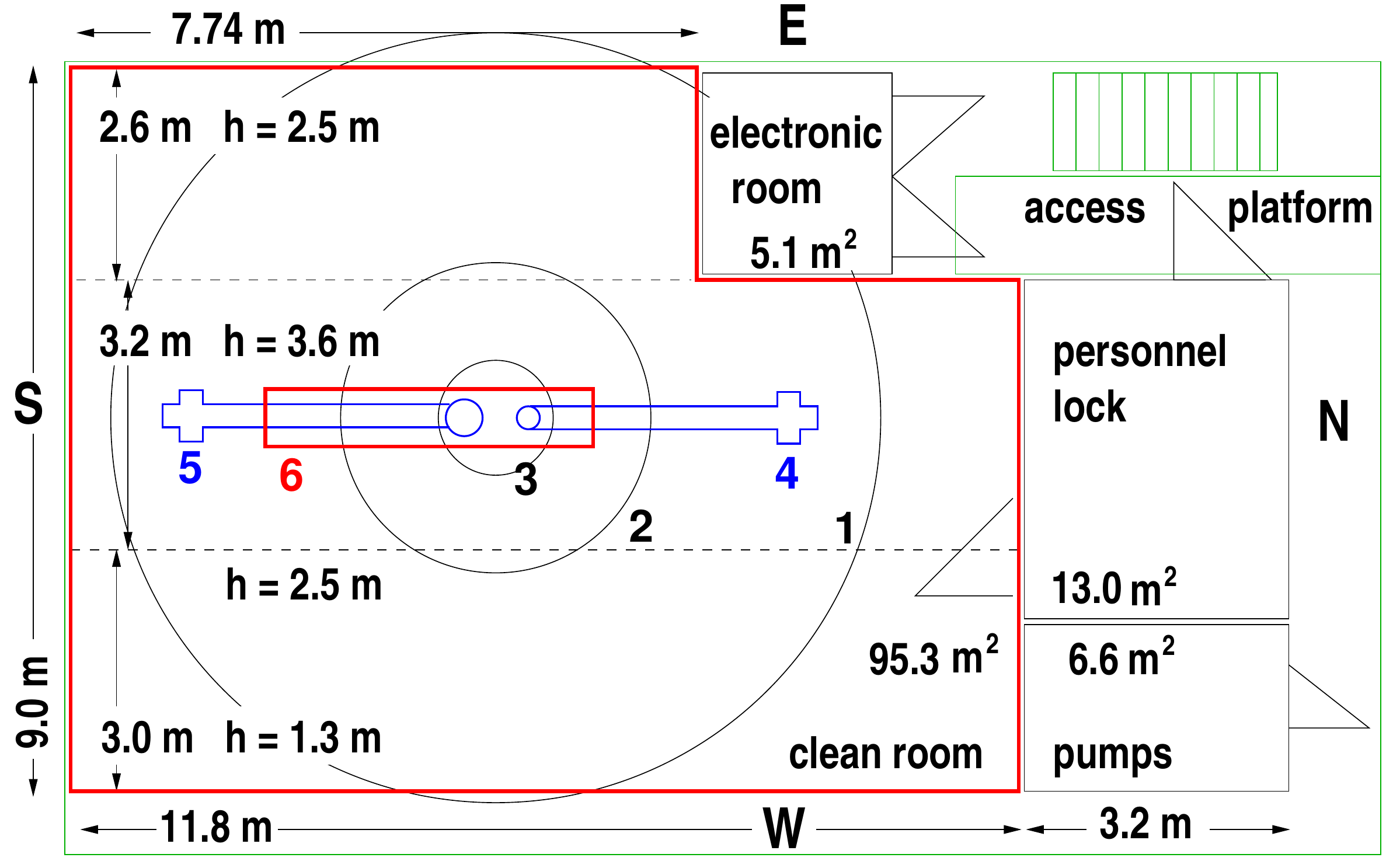}
\fi%
\caption{\label{fig:clean_room_layout}
    Plan of the platform on top of the \gerda\ building showing the clean
    room and auxiliary cabinets. The positions of the water tank (1), the
    cryostat (2) and its neck (3), all below the platform, are indicated.  The
    two arms of the lock and detector suspension system (4,5) are connected to
    the cryostat. The lock is enclosed by a glove box (6). The height of the
    clean room varies from 1.3 to 3.6~m.
}
\end{center}
\end{figure}

 The platform on top of the \gerda\ building supports the infrastructure for
 the clean handling and deployment of the Ge detectors into the cryostat
 without exposing them to air.  This infrastructure is designed as a gradient
 of radon reduction and cleanliness (Fig.~\ref{fig:clean_room_layout}). First
 a clean room is the working environment for experimenters within which a
 nitrogen flushed glove box is the working environment for the detectors. At
 the center a lock system provides a clean change between the environments of
 the glove box and the cryostat for detector insertion. The personnel lock and
 two small side rooms complete this complex.
%
\subsubsection{The clean room}
\label{sssec:cleanroom}

 The clean room is a class~7 room (ISO 14644-1~\cite{iest}) corresponding to
 less than ~10.000~particles/ft$^{3}$ of diameter $\geq$\,0.5~\mum. An
 overpressure of up to 30~Pa is maintained by pressing filtered air into the
 clean room.  The air volume of the clean room can be exchanged 49 times per
 hour.  Access to the clean room is via a personnel lock where an overpressure
 of 15~Pa is maintained.  The temperature inside the clean room is kept
 constant with variations of up to $\pm0.3$ degrees during normal operation.
 Maintaining the temperature within these bounds is required to prevent
 significant gain drifts in the electronics.  The relative humidity is
 regulated to (50$\pm$20)~\%. The constancy of these parameters depends to
 some extent on the LNGS cooling water supply of the underground laboratory.

 The ceiling of the clean room follows the curved shape of the ceiling of
 Hall~A, such that the central part of the clean room has a height of 3.6~m
 while the height at the wings reduces to a maximum of 2.5~m (see
 Fig.~\ref{fig:clean_room_layout}). The central part is equipped with two
 cranes at a height of 3.3~m that are movable along the south-north (S-N)
 direction, each with a maximum load of 500~kg. Both the southern wall and the
 central roof component are demountable. A maximum load of 150~kg/m$^2$ can be
 supported on the roof, greater than the load of the plastic muon veto system
 (sec.~\ref{ssec:muonveto}).  Adjacent to the clean room is an electronic
 cabinet with a cable tray feedthrough to the clean room.  Another adjacent
 room houses the pumps for the gas system of the lock.

 The class 7 specifications have been met during all times while the clean
 room was operating.  A LabView program monitors and outputs in a web interface
 the following parameters: particle measurements, radon content, overpressure,
 temperatures, and humidity.

\subsubsection{The twin lock and the suspension system}
 \label{sssec:lock}

\begin{figure*}[th!]
\begin{center}
\ifmakefigures%
     \includegraphics[width=\textwidth]{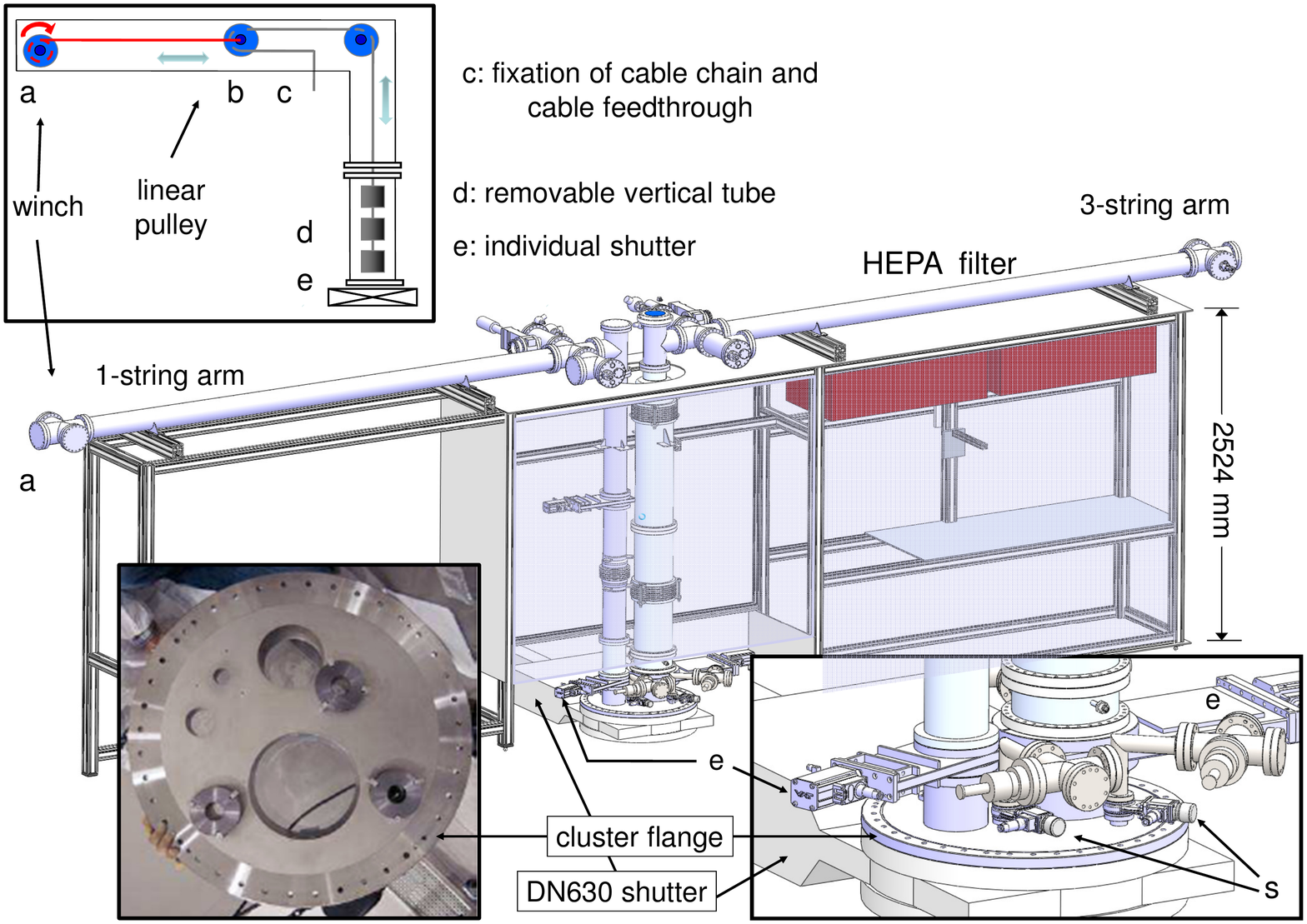}
\fi%
  \caption{\label{fig:CLock}
     Sketch of the twin lock for Phase~I with its two arms on top of the DN630
     shutter flange. The transparent blue area indicates the glove box with
     the HEPA filters (red). Each arm has an individual lock shutter (e) above
     which the vertical tube (d) can be removed to allow the insertion of the
     detector strings. The upper inset demonstrates the principle of the lock
     system: steel band (red) and cable chain with cables (black), winch (a),
     linear pulley (b), fixation of cable chain with cable feedthroughs (c),
     movable tube (d), and individual lock shutter (e).  The inset at the
     bottom right details the DN630 shutter, the cluster flange, the
     individual tube shutters and also two of the three source insertion
     systems (s) above the DN40 shutters. A picture of the bottom side of the
     cluster flange is shown in the bottom left. Visible are the DN160 and
     DN250 openings as well as the 3 smaller openings for the sources with the
     tantalum absorbers (and 2 spare holes).
}
\end{center}
\end{figure*}

 The twin lock system for Phase~I consists of two independent arms
 (Figs.~\ref{fig:clean_room_layout} and~\ref{fig:CLock}) that are connected
 with the cryostat via a cluster flange on top of the DN630 shutter (bottom
 inset of Fig.~\ref{fig:CLock}, see also sec.~\ref{ssec:cryo} and
 Fig.~\ref{fig:cryostat}).
 Inside each arm is a cable chain (top inset of Fig.~\ref{fig:CLock}), the
 mechanics for lowering the detector strings into the cryostat and lights and
 cameras for observation during this procedure.  One lock arm supports three
 detector strings inside a vertical tube of 250~mm diameter, while the other
 supports a single detector string inside a vertical tube of 160~mm diameter.
 Since the arms are part of the argon gas volume during data taking, they are
 built according to the European pressure vessel code.  The locks are
 constructed from stainless steel tubes that are connected either by welding
 or by CF metal seals.  The vertical section where the detector strings are
 mounted are both located inside a glove box where HEPA filters further reduce
 particle concentration. Each vertical part consists of two about 1~m long
 tubes which exhibit the functionality of an independent lock for one or three
 detector strings, respectively.

 Each lock arm may be closed from the cryostat by individual shutters (item~e
 in Fig.~\ref{fig:CLock}) allowing for the independent operation of each when
 the DN630 shutter is opened. The removal of the lower part of the vertical
 tube (item d) allows for the insertion of detector string(s) into the lock.
 The Ge diodes are transferred in evacuated containers into the glove box that
 is purged with boil-off nitrogen gas.  Within the glove box, germanium diodes
 and their front end electronics are assembled into strings of up to three
 diodes each (discussed in sec.~\ref{ssec:frontend}; a fully mounted string is
 shown in Fig.~\ref{fig:CC2location}).  These strings are then transferred
 into the lock. After the closure of the lock, it is evacuated and purged with
 argon gas.  The two lock volumes are connected individually to a pumping
 station and to the cryostat through a dedicated gas system.  The latter has
 been helium leak tested at a level of $10^{-6}$ ~(Pa~$\cdot\ell$)/s.

 As radon can diffuse through plastic, metal seals are used almost exclusively 
 for the lock system.  All non-metal materials were screened for radon
 emanation (see sec.~\ref{ssec:screening}, Table~\ref{tab:lock_rn_emanation}).
 The DN630 shutter is connected with Helicoflex metal seals, while a Kalrez
 seal is employed for the shutter itself.  The flange with the motor axle
 feedthrough has a double seal EPHD O-ring. To avoid radon diffusion through
 this non-metal seal, the feedthrough is constantly pumped. The leak rate of
 the motor connection was measured to be about 10$^{-5}$~(Pa$\cdot\ell$)/s.

\begin{table*}[t]
\begin{center}
\caption{\label{tab:cables}
        Cables deployed in the 1-string and 3-string locks.
}
\vspace*{2mm}
\begin{tabular}{lclcc}
cable & ref. &type &  1-string & 3-string \\
\hline 
Habia SM50                 &\cite{habia}& 50~$\Omega$, coaxial   & 15   & 24 \\
SAMI RG178                 &\cite{sami} & HV ~(4~kV), coaxial    & ~4   & -  \\
Teledyne Reynolds 167-2896 &\cite{teledyne}& HV (18~kV), coaxial & -    & 10 \\
Teledyne Reynolds 167-2896 &\cite{teledyne}& HV ~(5~kV), unshielded &~1 & ~2 \\
\hline
total number               &               &                     & 20 & 38\\
\end{tabular}
\end{center}
\end{table*}

 The scheme of the suspension system is shown in the top inset of
 Fig.~\ref{fig:CLock}.  The cable chain is fixed inside the lock (item c) and
 runs along the 3.6~m long horizontal tube. It is deflected at the far end of
 the tube by 180$^\circ$ around the ``linear pulley'' (item b), a pulley that
 is free to move in the horizontal direction by sliding on a linear
 bearing. Above the cryostat the chain is deflected by 90$^\circ$
 vertically. The linear pulley is connected to a metal band that rolls around
 a winch (item a) fixed to the axle of a stepper motor. By unrolling the metal
 band, the linear pulley moves towards the cryostat neck and the chain can be
 lowered into the cryostat.

\begin{figure}[bh]
\begin{center}
\ifmakefigures%
    \includegraphics[width=.8\columnwidth]{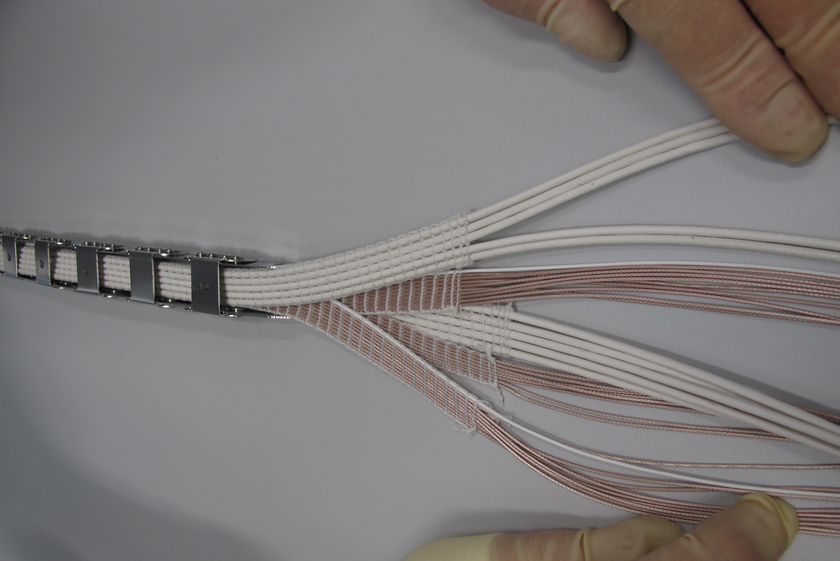} 
\fi%
  \caption{\label{fig:HandWovenCables}
    Woven cable bundles in the cable chain of the 3-string lock.
}
\end{center}
\end{figure}

 The cable chain supports the detectors mechanically and provides a conduit
 for the signal and high voltage cables to operate them. It is constructed
 from stainless steel that was selected for radiopurity. Its cross section is
 21$\times$13~mm$^2$ with a fillable area of
 17$\times$8~mm$^2$. Table~\ref{tab:cables} shows the configuration of the
 respective cable bundles for the 1- and 3-detector string case.
 In the 1-string bundle all cables are wrapped in a PTFE spiral coiled
 tube. This protects them against damage while moving inside the cable
 chain during its operations. The higher number of cables needed to operate
 nine detectors could be accommodated only by weaving the cables with PTFE
 thread into flat cables and protecting them against friction with the bottom
 of the cable chain by a thin metal band (see Fig.~\ref{fig:HandWovenCables}).

 The chain movement and the shutter status are controlled by a dedicated
 PLC. Inductive sensors are used as end switches. The position of the chain is
 determined redundantly by counting the number of turns of the motor and by a
 measuring tape with holes.  An optical system counts evenly spaced holes in a
 steel tape that is unrolled as the chain is lowered.  A friction clutch
 mounted between feedthrough and motor gear protects against excessive force
 transmission onto the cable chain.

\subsection{The calibration system}
 \label{ssec:calib}

 Regular calibration measurements with radioactive $\gamma$ sources provide the
 data necessary to determine the energy calibrations and resolutions of the
 diodes and to monitor their stability.  The energy scale is tracked via
 monitoring of specific $\gamma$ lines to identify
 periods in time for which single diodes showed a degraded performance.  These
 time periods can be identified and omitted or specially treated in the final
 analysis.

\begin{figure}[b]  
\begin{center}
\ifmakefigures%
	\includegraphics[width=\columnwidth]{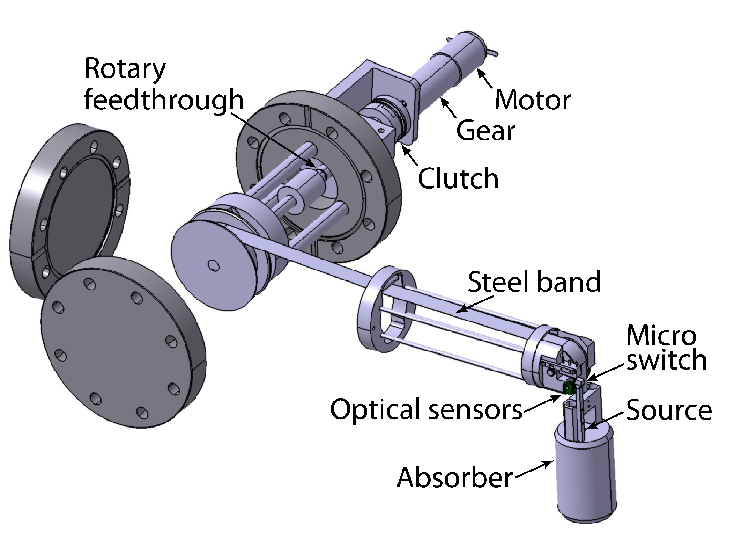} 
\fi%
\caption{\label{fig:calsys}
          A schematic view of one of three units of the calibration system
          (see also insets of Fig.~\ref{fig:CLock} and Ref.~\cite{phdfroborg}).
}
\end{center}
\end{figure}
 In order to calibrate the detectors within the LAr cryostat, three $^{228}$Th
 calibration sources are brought into the vicinity of the crystals. This is
 achieved by three vacuum sealed mechanical systems (Fig.~\ref{fig:calsys})
 that are mounted on top of the cluster flange (Fig.~\ref{fig:CLock}).  The
 systems can be individually decoupled from the cryostat via DN40 gate valves
 with electrical state indicators. To ensure that the background from the
 calibration sources is negligible during physics data taking, the sources are
 mounted on top of tantalum absorbers of 60~mm height and 32~mm diameter
 (Fig.~\ref{fig:calsys}). These movable absorbers rest inside the ones
 mounted on the cluster flange (left inset of Fig.~\ref{fig:CLock}).  Each
 absorber with its source is connected to a perforated stainless steel band
 which is deflected by $90^\circ$ before being rolled on spindles mounted
 behind horizontal band guides. The spindles are connected via magnetofluid
 sealed rotary feedthroughs to planetary geared DC motors.  Friction clutches
 between the feedthroughs and the motors protect against excessive force
 transmission on the steel band. The sources are moved with a speed of
 10~mm/s.

 Each unit has two redundant positioning systems. An incremental encoder
 counts the holes of the steel band that is perforated at 4~mm pitch.  The
 incremental encoder is mounted below the $90^\circ$ deflection point at the
 end of the band guide. At the same position a microswitch defining the null
 position is mounted.  A custom designed feedthrough, mounted on a CF flange,
 passes electronics for the incremental encoder as well as a gas line with a
 VCR 1/2'' gas connection.  The latter allows to pump and purge the source gas
 volume after the installation and before the shutter to the cryostat is
 opened.

 The second positioning system is based on a magnetic sampling multi-turn
 absolute encoder with 13 bit resolution, registering changes in position even
 if not powered.  The absolute encoder is mechanically coupled via a gear wheel
 to the external drive shaft on which the DC motor is mounted.

 The three source systems are controlled by a common control unit enabling the
 communication between a micro-controller and a PC via an RS422
 interface. Each calibration source can be individually moved via a control
 panel displaying the actual position and status of the respective unit.  The
 control panel also allows for a manual movement of the sources.  Correction
 functions for the thermal contraction of the steel band immersed in the
 cryo-liquid are applied when calculating the position. The incremental
 encoder serves as the main positioning system, while the absolute encoder is
 calibrated with respect to it.
 
 The RS422 interface allows to remotely control the system via a LabView
 GUI~\cite{labview}.  The GUI allows to automatize source movements, to change
 relevant settings, and to monitor the status of the sources and the
 controlling unit.  A closed or undefined gate valve state vetoes any motor
 activity on the LabView side. Malfunctions of the systems are monitored by
 the micro-controller that blocks any further movement of the sources in case
 an error occurs.

 Tests of the calibration systems prior to mounting on the cryostat showed an
 accuracy of the incremental encoder of $\pm2$~mm while the absolute encoder
 shows an accuracy of $\pm1$~mm. The position reproducibility is $\pm1$~mm.
 Tests using automatic sequencing proved that the long-term reliability of the
 systems is sufficient for their planned operation time.  The calibration
 system was installed in the \gerda\ cryostat in June 2011.  During the
 commissioning of a prototype system, a source dropped to the bottom of the
 cryostat due to the rupture of the supporting steel band. For Phase~I, the
 resulting contribution to the BI is negligible
 ($\le$\cpowten{1.1}{-3}~\ctsper\ based on $\sim$22~kBq in November 2011). The
 final version of the calibration system is working without any problems.

 The energy calibration of the diodes is performed by using 7 prominent lines
 in the $^{228}$Th spectrum: 510.8\,keV, 583.2\,keV, 727.3\,keV, 860.6\,keV,
 1620.5\,keV, 2103.5\,keV and 2614.5\,keV. For the calibration function a
 second order polynomial is used to account for ballistic defects of the
 measured pulses and for non-linearities of the electronics.  Calibration
 spectra with the resolutions of all detectors are shown in
 sec.~\ref{ssec:stability}.
%
\subsection{The Muon Veto}
\label{ssec:muonveto}

 The Gran Sasso overburden of 3500\,m\,w.e.~reduces the flux of cosmic muons
 to about 1.2\,/(h$\cdot$m$^2$) and shifts the mean energy to 270~GeV.  Muons
 penetrating the detector will lose energy by both electromagnetic
 interactions and by inelastic reactions with nuclei in which high energy
 neutrons can be produced. These neutrons will cause inelastic interactions
 themselves and produce more isotopes and neutrons. Hence muons are both a
 direct and indirect background source.

 The instrumentation of the water buffer surrounding the cryostat provides a
 cost-effective solution for the identification of muons by the detection of
 their Cherenkov light with PMTs.  Muons that enter the cryostat through the
 neck might only pass through a small water volume. An array of plastic
 scintillators on the roof of the clean room provides additional covering to
 detect muons passing this region. Signals from both detector systems are
 combined as a muon veto serving the germanium DAQ.  The muon veto system is
 designed to reduce the BI contribution from the direct muon events to a level
 of 10$^{-5}$~\ctsper\ at $Q_{\beta\beta}$ in the region of interest.

\subsubsection{The water Cherenkov detector}
         \label{sssec:cerenkov}
 MC simulations have been used to optimize the setup and in particular to
 define the number of photon detectors inside the water tank~\cite{phdKnapp}.
 A reflective foil glued on the walls of the water tank and cryostat
 contributes significantly to the light collection efficiency. This VM2000
 foil, produced by 3M~\cite{vm2000}, has a reflectivity of $>99~\%$ over a
 wavelength range of 400 to 775~nm and performs well as wavelength shifter for
 UV light that is re-emitted in the visible spectrum.  The foil has a rather
 small thickness of 206~\mum, i.e. $\sim$0.25~kg/m$^2$. Almost all outer
 surfaces of the cryostat, the inner wall and the floor of the water tank are
 covered with this foil.

\begin{figure}[bh!]
\begin{center}
\ifmakefigures%
  \includegraphics[width=0.78\columnwidth]{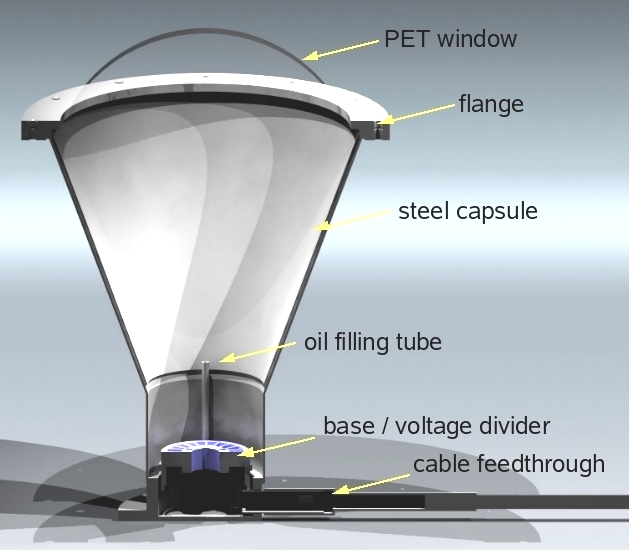}
\fi%
\caption{\label{fig:capsule}
    Schematic drawing of the encapsulation for the PMTs of the Cherenkov muon
    veto. The PMT, the oil and the silicone are not shown.}
\end{center}
\end{figure}

 For the Cherenkov light detection 8'' PMTs from ETL, type
 9350KB/9354KB~\cite{etl}, have been selected. They withstand an overpressure
 of 2$\cdot$\powten{5}~Pa absolute which is more than the pressure due to the
 maximum water height of 8.5~m. Since the PMTs are located outside the
 cryostat, there are no stringent constraints on their radioactivity.
 Nevertheless, 23 low-activity PMTs available were mounted in the almost
 closed water volume within the skirt of the cryostat (see
 Figs.~\ref{fig:overview} and~\ref{fig:cryostat}) and on the bottom plate of
 the water tank. The selected ETL 9454KB PMTs have e.g. a potassium content
 reduced from 300~ppm to 60~ppm.  All PMTs are encapsulated in stainless steel
 housings to prevent water from reaching the electrical contacts as shown in
 Fig.~\ref{fig:capsule}.
 In addition, the housing also acts as a mechanical support and mount for the
 PMTs. It consists of a steel cone, fixed to a bottom plate. To keep the PMT
 in position, the electrical base is fixed with polyurethane. As additional
 waterproofing, the electrical contacts are potted with silicone. The optical
 face of the housing is closed with a polyethylene terphthalate (PET) window.
 The volume between window and PMT is filled with mineral oil for a better
 matching of the refractive indices.
 
 The electrical power for a PMT and its signal readout is provided by a 
 single HV coaxial RG~213C/U cable with polyurethane cladding that is
 designed for underwater applications~\cite{jowo}. To facilitate timing, all
 cables have the same length of 35~m. In the electronics room, a splitter
 separates the HV and the signal lines.  Extensive tests have been performed
 to secure the underwater tightness of the capsules. One prototype was
 operated at full HV inside a pressure tank for several years without degraded
 performance. Independent long term tests of material degradation and cable
 performance were also performed.
    
 The water tank is equipped with 66~PMTs yielding a nominal coverage of
 0.5\%. Of these, 6 are mounted on the skirt facing inwards into the volume
 below the cryostat. Due to the few small openings
 (see. Fig.~\ref{fig:cryostat}), this part is an almost independent water
 volume separated from the main.  In the main volume, sets of 8 and 12 PMTs
 are mounted to the floor of the water tank facing upwards in rings of 5.5 m
 and 8.5~m diameter, respectively. The remaining 40 PMTs are mounted to the
 outer wall of the water tank facing inwards in four rings of 10 PMTs at the
 heights of 2~m, 3.5~m, 5~m, and 6.5~m.  The high voltage is supplied via a
 CAEN SV1527LC crate housing 6 CAEN A1733P high voltage cards~\cite{caen} with
 12 positive high voltage channels each.

 Five diffuser balls~\cite{Ritter:2010}, each equipped with a single LED, are
 distributed throughout the water tank. When pulsed they illuminate the entire
 water tank volume allowing for tests of all PMTs simultaneously. The PMT
 gain is adjusted and the calibration, in units of photo electrons (p.e.), is
 made using this system. In addition, each PMT can be triggered individually
 through an optical fiber for calibration, monitoring, and
 testing~\cite{phdRitter}. The initial HV was set for a gain of 2$\cdot$10$^7$
 for each PMT.

\subsubsection{The plastic scintillator muon veto}
    \label{sssec:plastic}
 The second part of the muon veto system consists of 36 plastic scintillator
 panels located on the roof of the \gerda\ clean room above the neck of the
 cryostat.  Each scintillator panel consists of a sheet of plastic
 scintillator UPS-923A~\cite{scint-ups} with dimensions of
 200$\times$50$\times$3~cm$^3$, an attached electronics board housing a PMT
 (one of 17 H6780-2~\cite{hamamatsu} or 19 PMT-85~\cite{kvadrotech}) and
 trigger electronics.  The light produced inside the plastic panel is guided
 to the PMT via 56 Y11~\cite{y11} optical fibers. They are glued to both of
 the 200$\times$3~cm$^2$ side areas of the panel.
 
 The 36 panels are stacked in three layers with 12 modules each, covering an
 area of 4$\times$3~m$^2$ in the N-S direction and centered at the neck of the
 cryostat. The panels in the second layer are placed directly on top of the
 first in the same orientation. The inner 8 modules of the third layer are
 rotated $90^\circ$ degrees with respect to the lower modules to create a
 finer pixelization.

\subsubsection{The trigger of the muon veto}
   \label{sssec:muondata}
 The data acquisition for the muon detectors is described in
 sec.~\ref{ssec:vmedaq}. For a valid trigger of the Cherenkov system, at least
 5~FADCs have to deliver a trigger signal within 60~ns. The threshold in each
 FADC channel is set such as to accept single photons with about 80\%
 efficiency.  The Cherenkov muon veto system is running smoothly since the
 beginning of the commissioning runs. Three out of the 66 PMTs in the water
 tank have been lost during two years of operation.

 A standard pulse from the Cherenkov detectors has a width of about 20~ns
 followed by a small overshoot and an electronic reflection around 350~ns
 after the main pulse.  The heights of each secondary pulse was less than 1/10
 of the main pulse causing no problem for the trigger system.  Pulse height
 calibration is employed to adjust the gain of the individual PMT. At
 sufficient low light pulser rates the PMTs can be set to the same response by
 adjusting slightly the respective HV. Since September 2010 the PMTs have been
 checked periodically for stability.  Only a few HV channels needed to be
 re-adjusted during that period.  The single photon peak is clearly
 distinguishable with a peak-to-valley ratio approaching 3.

 Within an event, the arrival time of pulses with a large light production is
 widely spread with differences up to 340~ns. Nevertheless, around half of the
 PMTs fire within the first 60~ns; therefore, this time interval has been
 chosen as coincidence time window for the trigger. The time spread is
 produced in part by the reflections on the VM2000 foil for the benefit of
 higher light yield.
 
\begin{figure}[b]
\begin{center}
\ifmakefigures%
  \includegraphics[width=.95\columnwidth]{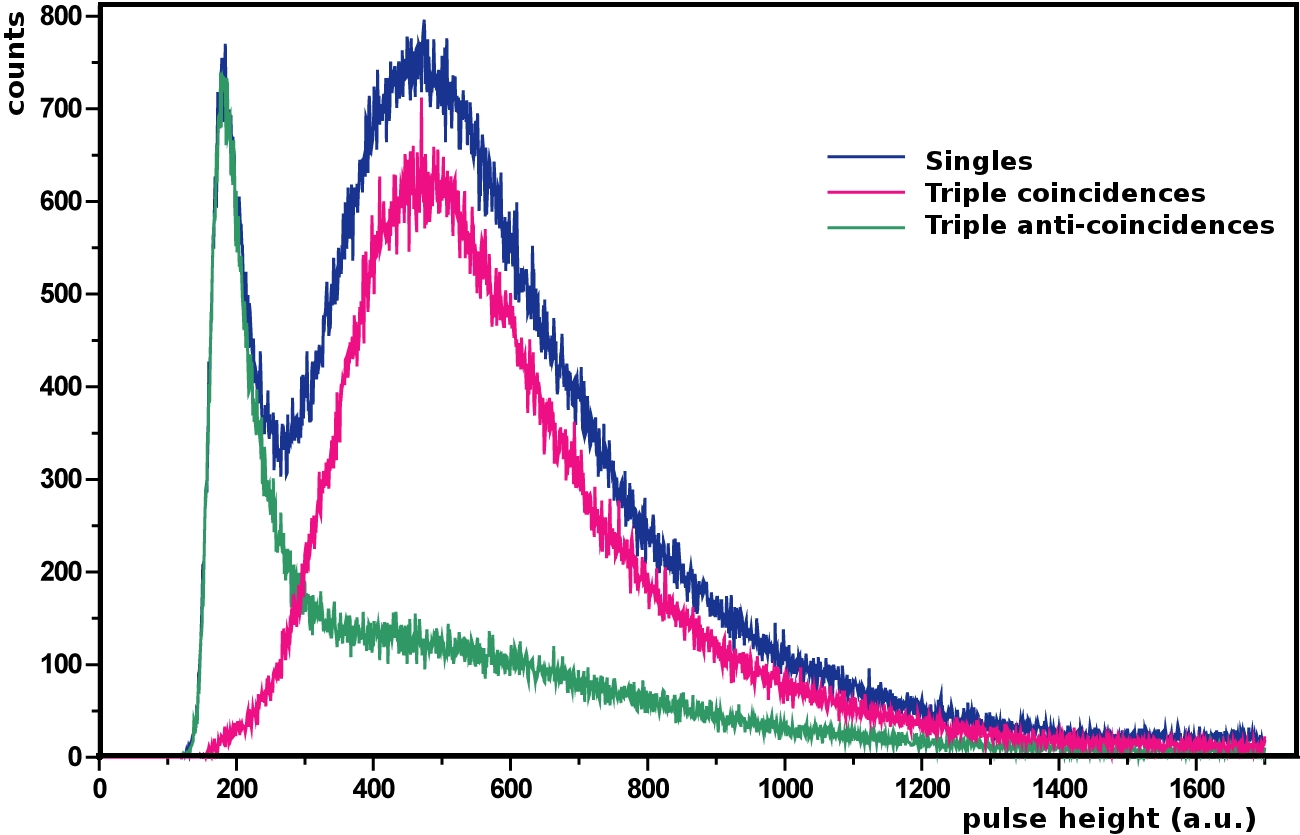}
\fi%
\caption{\label{fig:panelph}
       Spectra taken with the plastic panels: singles (blue), triple
       coincidences (pink), and their difference (green). 
  }
\end{center}
\end{figure}

 As to the plastic muon veto system, the triple coincidence between the layers
 allows for a clear separation of $\gamma$ background and muons. This is
 demonstrated by the spectra shown in Fig.~\ref{fig:panelph}.
  In the singles
 spectrum (blue), the low energy part due to $\gamma$ rays is dominating and
 it exhibits a long tail to higher energies. The triple coincidences reveal
 unambiguously the minimum bias signal of muons.

 The triggers of the Cherenkov and the plastic detector systems are combined
 via a logic OR that is recorded by the germanium DAQ. 

\section{Readout, Data Acquisition and Processing}
  \label{sec:data}
\subsection{The front-end electronics} 
 \label{ssec:frontend}

\begin{figure}[b]
\begin{center}
\ifmakefigures%
    \includegraphics[width=\columnwidth]{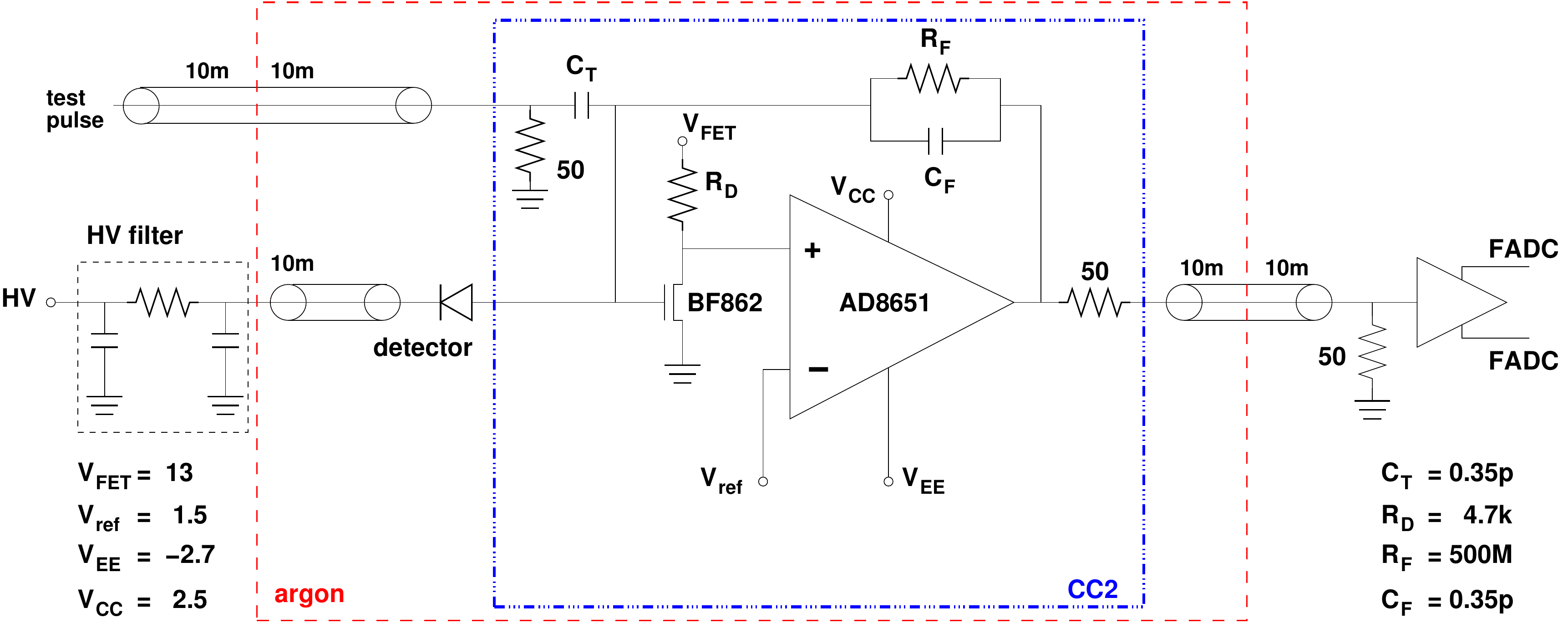}
\fi%
\caption{\label{fig:CC2schematic}
    The scheme of the \gerda\ front end circuit including grounding and
    cable lengths. The parts within the dotted box are on the CC2 PCB.
    The red dashed line shows the limits of the argon volume. The resistor
    values are given at room temperature.
 }
\end{center}
\end{figure}

\begin{figure}[t]
\begin{center}
\ifmakefigures%
      \includegraphics[width=\columnwidth]{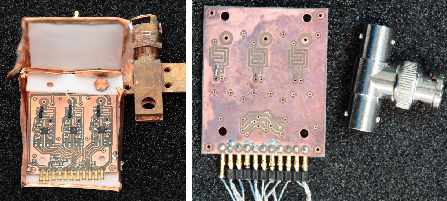}  
\fi%
\caption{\label{fig:CC2lowactivity}
              The CC2 PCB front and back side integrating three channels.
}
\end{center}
\end{figure}

 Germanium detectors are normally read out with charge sensitive preamplifiers
 (CSP). In commercial devices the input transistor, a JFET, and the feedback
 components are close to the diode and the JFET is cooled to about 100~K for
 optimal noise performance.  The rest of the CSP is at ambient temperature at
 a small distance of typically 50~cm.  In \gerda, the same scheme would result
 in a distance of 10~m between the cold and the warm part of the CSP. The
 signal propagation time to close the feedback loop would consequently be
 longer than 100~ns. This would limit the bandwidth or lead to oscillations
 and the pulse shape information would be largely lost.  To avoid this loss we
 operate the entire CSP in LAr. The minimal allowed distance between the
 detectors and the preamplifier depends on the radioactivity of the
 latter. The schematic of the implemented CSP (called CC2~\cite{cc2reference})
 is shown in Fig.~\ref{fig:CC2schematic}.
 The input JFET is a BF862 from NXP Semiconductors and the second stage is the
 AD8651 from Analog Devices. Both components are used in commercial packages.
 Three channels are integrated on a single layer Cuflon PCB (delivered by
 Polyflon~\cite{polyflon}).  The feedback and test pulse capacitors are
 implemented as stray capacitances between traces on the PCB board (see
 Fig.~\ref{fig:CC2lowactivity}).  Tantalum filter capacitors are used solely
 and a separate line driver is omitted to limit the radioactivity (see
 sect.~\ref{sssec:o_screen}).

\begin{figure}[b]
\begin{center}
\ifmakefigures%
   \includegraphics[width=0.95\columnwidth]{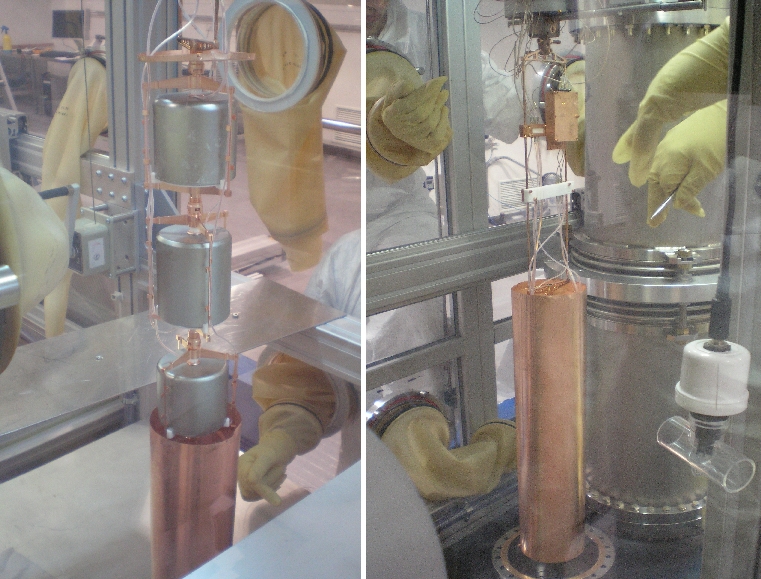}
\fi%
\caption{\label{fig:CC2location}
     Left: a string of three \geenr\ detectors is inserted into the mini-shroud.
           This work is performed in the glove box of the clean room.
     Right: closed detector string and 3-channel CC2 preamplifier inside
            a copper box about 30~cm above the string. The connections between
            CC2 and detectors are made with Teflon insulated copper strips
            that are tightly fixed to prevent microphonics.
            In the background, part of the 3-string lock is visible.
}
\end{center}
\end{figure}

 The CC2 is located inside a copper box (Fig.~\ref{fig:CC2location}, right)
 that provides electromagnetic shielding.  The input wires connecting to the
 detectors are copper strips with 2$\times$0.4~mm$^2$ cross section produced
 by wire erosion from screened material. The insulator is a PTFE tube. The
 same scheme is used for the last part connecting the high voltage cable to
 the detector.  All copper strips are fixed along the detector supports to
 avoid microphonics.

 Realizing the CC2 with the values given in Fig.~\ref{fig:CC2schematic}, its
 specifications are: sensitivity of 180~mV/MeV, input range at least 10~MeV,
 power consumption of less than 45~mW/channel, cross talk between channels of
 less than 0.1~\%, rise time with terminated analog output of typically 55~ns,
 decay time of 150~\mus.  The noise (converted to energy equivalent FWHM for
 Ge) is typically 0.8~keV + 0.024 keV/pF for a 10~\mus\ semi-Gaussian pulse
 shape with 8~\% systematic uncertainty attached.  For a 600~g coaxial
 detector the energy resolution achieved was 1.96~keV at the 1274~keV $\gamma$
 line of $^{22}$Na.

 A pulser signal is sent periodically to the test pulse input of the CC2
 (Fig.~\ref{fig:CC2schematic}). The voltage step at the capacitor $C_T$
 injects a fixed charge at the input of the CC2 and thus allows a monitoring
 of the entire readout chain during data taking.

 Thin coaxial cables from Habia (type SM50, 94~pF/m, 0.9~$\Omega$/m,
 \cite{habia}) are used for the analog outputs of the CC2 as well as for power
 supply (see Table~\ref{tab:cables}). Welded BNC feedthroughs act as seal
 between air and the cryostat/lock.  The ground is connected via the lock with
 the cryostat and water tank.  RG178 cables transmit signals from the BNC
 connectors to the FADC in the electronics cabinet where the analog signals
 are digitized. The total cable length that the CC2 must drive the signal over
 amounts to 20~m.

 The HV feedthroughs between air and the argon gas inside the lock are custom
 made. For leak tightness the braid of the HV cable is replaced for a few
 centimeters by a solid wire. The latter together with the soldered connection
 and part of the cable is then encapsulated with epoxy (Stycast (R)
 FT2850,~\cite{stycast}) inside a 5~cm long stainless steel pipe with a CF16
 flange at one end. This solution avoids discharges in the argon gas with up
 to 6~kV bias on the cable.  The HV cable shielding is connected to this pipe
 and thus also to the lock. At the air side of the feedthrough, a filter is
 mounted to reduce electromagnetic noise.
 
 The HV cables inside the lock are Sami RG178~\cite{sami} and Teledyne
 Reynolds 167-2896~\cite{teledyne}. They end about 30~cm above the top
 detector from where the above mentioned copper strips are used. The HV bias
 is provided by NIM modules from CAEN (N1471H, 4ch Power Supply, \cite{caen}).

\subsection{The data acquisition}

 The data from the Ge detectors and from the muon veto system are acquired
 with two different data acquisition systems. Both systems are synchronized by
 a common  GPS pulse per second (PPS) signal. All signals are digitized by
 FADCs and the energy is reconstructed offline.

\subsubsection{The germanium DAQ for Phase~I}
    \label{ssec:customdaq}
 The custom made Phase~I DAQ~\cite{bib:customdaq,bib:customdaq_ur} for the
 germanium readout consists of NIM modules, PCI based readout boards and
 external logic for the trigger generation.  Each NIM module digitizes 4
 channels.  It accepts both single-ended and differential signals. The signal
 polarity as well as signal attenuation (0~dB/12~dB) or gain (0-6-12-18~dB) of
 the analog input stage can be selected via jumpers. The offset is adjustable
 and the bandwidth of the anti-aliasing filter is 30~MHz. Analog-to-digital
 conversion is based on the Analog Devices AD6645 A/D converter (14~bits,
 100~Msamples/s).  The digitized data are processed with a trapezoidal filter
 with programmable threshold to generate a trigger, averaged to monitor the
 baseline and sent via an LVDS link to a PCI board.  The synchronization of
 the different NIM modules is ensured by a common external clock and BUSY
 signal. The latter blocks the incoming data during readout.

 The custom-made PCI readout boards are mounted in a personal computer running
 the Linux operating system and are operated at 32~bit/33~MHz allowing for a
 maximum transfer rate of 132~MB/s.  Selecting a 40~\mus\ window, four of the
 10~ns samples are added (i.e.~the sampling rate is reduced from 100 to
 25~MHz) to reduce the data rate and readout time. For pulse shape analysis a
 5~\mus\ trace around the rising edge of the signal is stored in addition at
 full sampling rate.  The length allows for a 10~\mus\ shaping time for the
 moving window deconvolution algorithm\cite{stein}, and thus no information
 for the energy reconstruction is lost in the compression.  A 32~bit trigger
 counter and a 64~bit 100~MHz timestamp are saved together with the data. NIM
 logic builds the OR of all triggers and generates the BUSY signal.

 A Qt based~\cite{bib:qt} comprehensive graphical user interface for the whole
 system was implemented.  A JAVA-based Graphical Analysis tool was developed
 for online monitoring of the data.  Test measurements were performed with a
 BEGe detector.  The energy resolution was similar to the one obtained with
 spectroscopy amplifiers.

 In addition, a copy of the hardware used for the digitization and triggering
 of the muon veto signals (see below) is available for the readout of the
 germanium detectors.  Some parameters of the commercial FADCs, e.g. the
 shaping of the signal for the trigger, are adjusted to the preamplifier pulse
 shapes and the trace length is set to 160~\mus.  Both germanium DAQ systems
 are operated in parallel.

\subsubsection{The muon DAQ}
   \label{ssec:vmedaq}

 The muon data acquisition is installed in a VME crate housing 14 FADCs of
 type SIS 3301 from Struck (8~channels, 100~MHz, 14~bit,~\cite{struck}).  Each
 card has 2 memory banks of 128k samples which are divided into 4k size per
 event.  If one bank is full, writing continues to the second bank while the
 first one can be read out. This reduces the dead time to less than 0.1\%
 during normal data taking.  Each channel is equipped with an analog
 anti-aliasing filter with 30~MHz bandwidth and with a trapezoidal filter for
 the digitized data. If the filter output is above threshold a trigger is
 generated.  The logical OR of all 8 triggers in a FADC module is fed
 into a custom made VME board, called MPIC.  The MPIC generates a global
 trigger if a programmable number of cards output a trigger within a
 coincidence time window. This card also provides a time stamp for the event
 that is synchronized to the GPS PPS signal with 10~ns precision.  If a
 trigger occurs, a ``stop pulse'' is fanned out to all FADC cards to stop
 writing to the circular event buffer such that the data is saved for
 readout. Upon a trigger, 4~\mus\ traces for all channels are stored on disk.
 The stop pulse is also digitized as an additional analog input by the
 germanium DAQ to easily veto coincidence events.  Delayed coincidences can be
 detected by comparing the event time stamps between the muon and germanium
 events in the offline analysis.
 
 The muon veto calibration mentioned in sec.~\ref{sssec:muondata} is performed
 by powering five LED drivers with a digital-to-analog converter
 (PAS9817/AO~\cite{pasdac}) in connection with a CAEN V976~\cite{caen} fan-out
 for pulser signals. The light from the LEDs is uniformly distributed to all
 the PMTs through five diffuser balls.

\subsection{Data handling}
\subsubsection{Data flow and blinding}
    \label{ssec:dataflow}

 The binary raw data format is defined by the different data acquisition
 systems.  In order to optimize the analysis streaming and to provide a unique
 input interface for the analysis, all raw data are converted to a common
 standardized format. \mgdo\ (\majorana-\gerda\ Data Objects)~\cite{mgdo} is a
 software library jointly developed by \gerda\ and \majorana, that contains
 general-purpose interfaces and analysis tools to support the digital
 processing of experimental or simulated signals.  The custom data objects
 available in the \mgdo\ package are used as reference format to store events,
 waveforms, and other DAQ data (time stamps, flags).  The \mgdo\ data objects
 are stored as \rootv\ files~\cite{root}. The set of \rootv\ files produced by
 the conversion of raw data is named \tier1.

 Since the information contained in the \tier1 set and in the raw data is
 expected to be equal, the conversion procedure is the optimal place where
 blinding can be applied. Events with an energy close to $Q_{\beta\beta}$ are
 not exported to \tier1 but they remain saved in the backup of the raw data.

 The software framework \gelatio~\cite{gelatio} contains nearly independent
 and customizable modules that are applied to the input \tier1 waveforms. The
 results (pulse amplitude, rise time, average baseline, etc.)  are stored as a
 new \rootv\ file (\tier2).  A description of the analysis modules is
 presented in Ref.~\cite{pipeline}. Higher level \tier$i$ files can be created
 that contain additional parameters evaluated from more advanced analysis
 (e.g. calibrated energy spectra). The information of the same event stored in
 different \tier$i$ files can be accessed by means of the \rootv\ friendship
 mechanism~\cite{root}.

\subsubsection{Data storage}
   \label{ssec:datastorage}

 The data acquisition systems store data underground on a server with 14~TB
 space. Every night, the newly accumulated data are transferred to a
 \gerda\ server in the LNGS computing center that has 36~TB disk space. This
 server is only accessible for a small number of users such that the raw data
 are hidden and only blinded data are available for analysis.  Copies of the
 raw data are stored at LNGS, in Heidelberg, and Moscow.

\subsubsection{Quality control}
 \label{ssec:quality}

 The event reconstruction of new data occurs automatically once per day. Since
 our rate is low, it is possible to store filtered information like the event
 energy, pulse rise time or baseline level in a data base.  An interface
 allows simple access with a web browser or, alternatively, by a user written
 C++ program~\cite{quality}.  The event traces stored in \tier1 files can also
 be viewed.

 A list of predefined scripts generate monitoring histograms like trigger
 rates versus time, pulser stability plots, baseline shifts and energy
 distributions and these are checked daily.
\subsection{The \gerda\ network structure}
      \label{ssec:gerdanet}
 \gerda\ has a dedicated network in Hall~A.  It is connected to the external
 laboratories above ground by two dedicated multi-modal optical fibers.  They
 connect to a network switch~\cite{bib:2810-48G} that offers access security
 and advanced prioritization and traffic-monitoring capabilities. The
 different network lines are routed inside the \gerda\ infrastructure.

 The switch is directly connected to a dedicated ser\-ver~\cite{bib:r300} that
 provides network routing facilities and acts as a firewall and user
 authentication server.  At the moment, this is the only public service
 available directly from external networks and it is used to access all
 \gerda\ internal network resources and services.  A Port Address Translation
 (PAT) network device is used internally, to translate TCP/UDP communications
 between \gerda\ private network computers and public network hosts.

The following centralized services are available:
($i$)   {\tt NIS}-server for user authentication,
($ii$)  {\tt DNS}-server for host name resolution,
($iii$) {\tt DHCP}-server for the DAQ/slow control machines and
                 all the computers attached temporarily to the network
                 (i.e. laptops), and
($iv$)  {\tt Web}-server for the whole experiment.

 In order to provide access to internal \gerda\ resources (mainly internal Web
 servers), a {\tt proxy} service has been setup. Thus, it is possible to
 access internal Web servers through the main \gerda\ Web server.

\subsection{The slow control}
  \label{ssec:slow}

The \gerda\ slow control system~\cite{slowcontrol} is responsible for:
\begin{enumerate}   
\itemsep=0pt\parsep=0pt\parskip=1pt
\item monitoring of parameters characterizing the status of the subsystems
  (temperature, pressure, detector currents, etc.),
\item control and monitoring of low and high voltage power supplies through
  a graphical user interface (GUI),
\item  storage of the monitored values in a database  for later retrieval;
\item alarm handling,
\item web pages for subsystems breakdown,
\item online histograms for the relevant parameters,
\item reliable remote monitoring of the whole experiment.
\end{enumerate}
 The slow control consists of four building blocks.  A database is the core of
 the slow control system. It stores both the data and the configurations.
 PostgreSQL~\cite{postgresql} was selected as relational database SQL
 compliant with the capability of an embedded procedural language (PL). In
 case of a high number of records in the data tables, the database will be
 split in two: a so-called online database where all the data up to one week
 are stored and the historical database where older data are copied regularly
 after data compression.

 The acquisition task is performed by a pool of clients each serving a
 dedicated hardware subcomponent.  The clients store the acquired data in the
 database. Depending on the specific hardware, different types of connections
 are used by the clients: web access, CANbus, serial RS232, etc.  All data
 written into the database have a proper time stamp, that constitutes the main
 method to study correlations.  All hardware settings are stored in the
 database.

 Alarms generated automatically by some components go directly to
 the LNGS safety system and to the \gerda\ on call experts and  the slow
 control system will record the event into the database for future analysis.
 The alarm manager is a supervisor process that retrieves data from the
 database and is able to generate warnings or error messages in case of a
 malfunctioning sensor.

 The system is completed by a web interface where alarms, instant and
 historical data (through histograms) and the status of the clients can be
 seen.  The control interface is based on HTML.  The data are updated
 automatically using Ajax~\cite{ajax} in pull manner.

 The database has been operational since autumn 2009. Data collected in two
 and a half years are only 94 MB. This is in part due to the use of a data
 reduction policy at the level of the readout of some subcomponents (cryostat,
 clean room, water loop).


\section{Summary of screening results}
   \label{ssec:screening}

 A very careful selection of materials is critical to achieve our goal of one
 to two orders of magnitude reduction in backgrounds relative to previous
 experiments. For Phase~I, this selection was carried out by using
 state-of-the-art screening techniques during the design and construction
 phases. The screening facilities continue to be used in the preparations for
 Phase~II of \gerda.

 Material screening was performed mainly with the following three techniques:%
\begin{enumerate}
\item Gamma ray spectroscopy with High Purity Germanium spectrometers in four
  underground laboratories: at the Max-Planck-Institut f\"ur Kernphysik (MPIK)
  in Germany, HADES (IRMM) in Belgium, the Baksan Neutrino Observatory (BNO
  INR RAS) in Russia and at LNGS in Italy.  The ultimate detection limit for
  the best spectrometers in deep underground laboratories lies around
  10~$\upmu$Bq/kg for \Ra\ and \Th~\cite{Gempi1,Gempi2}.
\item Gas counting with ultra-low background proportional counters. They were
  originally developed at MPIK for the \Gallex\ solar neutrino
  experiment~\cite{Wink} and are used in \gerda\ for \Rn\ measurements.
\item Mass spectrometry with Inductively Coupled Plasma Mass Spectrometers
  (ICP-MS). The \gerda\ collaboration has access to two ICP-MS machines, one
  at LNGS and one at INR RAS, Moscow.
\end{enumerate}
 In addition, some dedicated samples were analyzed with Neutron Activation
 Analysis (NAA) and Atomic Absorption Spectroscopy (AAS).

 Altogether almost 250 samples were screened by gamma ray spectroscopy. The
 main focus was on electronics components (about 85 samples), metal samples
 (about 65 samples, mostly stainless steel and copper) and plastic materials
 (about 50 samples). Also the Rn emanation technique was extensively applied
 (about 120 samples) and about 20 samples were screened by ICP-MS. In this
 section some selected results, most relevant to the construction, will be
 given. Some more results can be found in
 Refs.~\cite{steel_paper,gamma_screening_gerda,rn_emanation_gerda}.

\subsection{Argon purity}
   \label{sssec:argon}

 The $^{222}$Rn concentration of commercial liquid nitrogen was
 measured~\cite{morex} and its purification to a level of 1~\mubq/m$^3$ at
 standard temperature and pressure has been demonstrated in the past for
 \borexino~\cite{Heusser_N2}.  For \GERDA\ the same questions arose for liquid
 argon since the $^{222}$Rn concentration in freshly produced argon was found
 to be in the range of mBq/m$^3$ (STP) which is about an order of magnitude
 higher than for nitrogen.  While this is not so relevant for the first
 cryostat filling, a constant refilling was considered for the case that the
 active cooling would fail (see sec.~\ref{sssec:cryostat}).

 Argon purification tests based on radon adsorption on low temperature
 activated carbon traps were performed with gaseous and liquid argon.  For the
 gas phase, reduction factors of more than 1000 were achieved for a 150~g
 trap~\cite{LAr-paper}.  These are similar to the results achieved for
 nitrogen~\cite{Heusser_N2}. For the liquid phase, in most cases a reduction
 factor of 10 could be achieved for a small 60~g column. In \GERDA\ an
 activated carbon column ($\sim$1~kg) was installed, which is expected to
 reduce the $^{222}$Rn concentration by two orders of magnitude.  All
 measurements were performed with the Mobile Radon Extraction unit
 (MoREx, \cite{Heusser_N2}).

\subsection{Radiopurity of the cryostat}
    \label{sssec:rn_cryo}

 Besides the argon, the second largest mass item in close contact to the
 diodes is the cryostat. It is made of austenitic stainless steel with an
 additional inner copper shield (see Figs.~\ref{fig:overview}
 and~\ref{fig:cryostat}). The stainless steel was procured in more than 10
 relatively small batches of a few tons and roughly a 50~kg sample of each
 batch was screened with gamma ray spectrometers~\cite{steel_paper}. During
 this campaign it was discovered that stainless steel may have low
 \Th\ activity that is about 10 times lower than what was known from earlier
 screening campaigns~\cite{BX_lowbackground}.  Finally, the cylindrical part
 of the cryostat, closest to the diodes, could be constructed from stainless
 steel batches with a \Th\ concentration below 1~mBq/kg. All other batches
 have a \Th\ concentration below 5~mBq/kg. Another contamination in stainless
 steel is \Co. In the batches for \gerda, a mean \Co\ activity of 19~mBq/kg
 was found~\cite{steel_paper}. The availability of low radioactivity stainless
 steel led to a significant reduction in the necessary mass of the inner
 copper shield.

\begin{table*}[t]
\begin{center}
\caption{\label{tab:rn_cryostat}
   Measurements on \Rn\ emanation of the \gerda\ cryostat at room temperature
   after various stages of construction.
  }
\vspace*{2mm}
\begin{tabular}{ccc|r@{ $\pm$ }l}
no. & date & description & \multicolumn{2}{c}{result [mBq]} \\ \hline
1 & Nov 2007 & after construction and first cleaning &
                                                  \hspace{0.5ex} 23.3 & 3.6 \\
2 & Mar 2008 & after additional cleaning                       & 13.7 & 1.9 \\
3 & Jun 2008 & after copper mounting                           & 34.4 & 6.0 \\
4 & Nov 2008 & after wiping of inner surfaces                  & 30.6 & 2.4 \\
5 & Sep 2009 & in final configuration                          & 54.7 & 3.5 \\
\end{tabular}
\end{center}
\end{table*}

 Any \Rn\ released from the inner surface of the cryostat will be dissolved in
 the liquid argon and may be transported to the germanium diodes by
 convection. Therefore, the \Rn\ emanation rate of the cryostat was measured
 after its construction. The cryostat was sealed, evacuated and filled with
 \Rn-free nitrogen gas that was produced with MoREx.  After a certain time in
 which \Rn\ could accumulate, the nitrogen was agitated (to assure a
 homogeneous radon distribution) and a sample of a few cubic meters nitrogen
 was extracted. Then the \Rn\ concentration in this aliquot was measured with
 low background proportional counters and the result was scaled to calculate
 the \Rn\ emanation rate of the entire cryostat. The measurement was repeated
 after various modifications of the cryostat and the results are summarized in
 Table~\ref{tab:rn_cryostat}.

 The first two measurements were performed when the cryostat was still empty,
 i.e. just the surface of the stainless steel vessel was under
 investigation. The cleaning then performed was a pickling and passivation
 treatment with an acidic gel.  In the first measurement a \Rn\ emanation rate
 of 23~mBq was measured. This reduced by a second cleaning cycle to a level of
 about 14~mBq. After the copper shield was installed a subsequent measurement
 showed an increase of the \Rn\ emanation rate by about 20~mBq. A plausible
 hypothesis was that dust was introduced during the copper mounting. However,
 this explanation was rejected because thorough surface wiping did not improve
 the result significantly (see measurement No.~4).

 The final configuration of the cryostat includes a manifold through which all
 tubing is distributed, a compensator to connect it to the lock, a radon
 shroud (see sec.~\ref{sec:det})  and many
 sensors and safety devices. The \Rn\ emanation rate of the cryostat in its
 final configuration is $(54.7\pm3.5)$~mBq. Assuming a homogeneous
 distribution of \Rn\ in the liquid argon, this would result in a contribution
 to the BI at \qbb\ of 7$\cdot$\vctsper.  To reduce this background, a
 cylinder made from 30~$\mu$m thick copper foil (called radon shroud, see
 item~4 in Fig.~\ref{fig:cryostat}) was installed around the diodes with the
 intention that \Rn\ that is emanated from the walls is kept at sufficient
 distance from the diodes.

\subsection{Radon emanation of components inside the lock}
  \label{sssec:radonlock}

\begin{table*}[H!t]
\begin{center}
\caption{\label{tab:lock_rn_emanation}
    Radon emanation of non-metallic materials used in the lock. The amount of
    the material used and the corresponding emanation is listed.  Values
    indicated by * are estimated by the detection of $^{226}$Ra using 
    $\gamma$-ray screening. They are conservative upper limits since not all
    $^{222}$Rn will escape the material.
}
\vspace*{2mm}
\begin{tabular}{lllr}
component                & amount & material &  total Rn\\
                         &        &          &  emanation rate\\
\hline
\up%
LED                      & 4 pieces        &         & (207$\pm$25)~\mubq\\
Kappa camera             & 4 pieces        &         &       $<$350~\mubq\\
inductive end switch     & 4 pieces   & mostly steel &  (73$\pm$13)~\mubq\\
meter drive head         & 2 pieces        &         & (860$\pm$180)~\mubq\\
meter drive plug         & 2 pieces        &         & (400$\pm$180)~\mubq\\
pulley bearings          & 12 pieces       & Iglidur &        $<$7.2~\mubq\\
linear pulley guides     & 4 pieces        & Iglidur &        $<$4.8~\mubq\\
O-ring seal shutter      & 1 piece         & Kalrez  & (400$\pm$100)~\mubq\\
O-ring motor feedthrough & 2 pieces        & EPDM    & (7.8$\pm$1.0)~\mubq\\
HV cables SAMI RG178     & 40~m (300~g) &            &        $<$680~\mubq \\
signal cables Habia SM50 & 508~m        &            & (273$\pm$50 )~\mubq\\
LV supply TR 5 kV        & 62~m&                     & (50.4$\pm$14.7)~\mubq\\ 
Thermovac pressure gauge
 \cite{thermovac}        & 2 pieces     &            &       $<$12.6~\mubq\\
BD diff. pressure sensor
 \cite{bdsensor}         & 3 pieces     &            & (117$\pm$18 )~\mubq\\
 \hline
\up
HV cables  TR 18 kV      & 1.53~kg (100~m)  &        &* $<$11.1~mBq\\
spiral coiled tube       & 273~g (11.2~m)   & PTFE   &* $<$3.2~mBq\\
\end{tabular}
\end{center}
\end{table*}

 The lock system is directly connected to the \gerda\ cryostat (see
 Fig.~\ref{fig:cryostat} and sec.~\ref{sssec:lock}).  Thus, \Rn\ that is
 emanated inside the lock may be dissolved in the liquid argon and can
 contribute to the background.  Consequently, the selection of low-emanating
 construction materials for the lock and items inside the lock was a rigorous
 process.  Flanges to the outside were sealed with metal gaskets whenever
 possible.  At places where O-rings had to be used Kalrez~\cite{kalrez}
 O-rings were chosen to avoid VITON, which is known to be a relatively strong
 source of radon.  The \Rn\ emanation rates of all Kalrez O-rings that are used
 in the lock system were investigated and it was confirmed that they are much
 radiopurer than VITON.  As a result of these measurements, an upper limit of
 0.6~mBq can be given for the integrated \Rn\ emanation rate of the subset of
 O-rings that are in direct contact with the inner volume of the lock.

 Table~\ref{tab:lock_rn_emanation} summarizes the results of all the other
 components in the lock that were screened for their \Rn\ emanation rate.  As
 can be seen in the right column the integrated radon emanation rate of all
 components is less than 17~mBq.  This is low compared to the \Rn\ emanation
 rate of the cryostat.  Moreover, there are cold copper surfaces in the argon
 gas phase just above the liquid level which will act as a getter. Therefore,
 the \Rn\ emanation of the lock is a minor source of background for \gerda.

\subsection{Further selected screening results}
\label{sssec:o_screen}

 Before the construction of \gerda\ it was already known that high purity
 copper is one of the most radiopure materials~\cite{Gempi2}. Therefore, it
 was the natural candidate material for the construction of the low mass diode
 holder (see Table~\ref{tab:thorium}).  As insulating material, one of the
 most promising candidates from previous measurements is PTFE. A batch of
 extruded PTFE was purchased that was produced under particularly clean
 conditions and screened with the
 \GeMPI\ spectrometer~\cite{gamma_screening_gerda}.

\begin{table*}[t]
\begin{center}
\caption{\label{tab:thorium} 
  Gamma ray screening results for selected materials. Given are 
  1$\sigma$-boundaries or 90~\% limits.  Note, one PCB board serves three
  detectors.
}
\vspace*{2mm}
\begin{tabular}{lrccc}
component              & amount    &$^{40}$K &  $^{226}$Rn    & $^{228}$Th\\
                       &        &[mBq/kg]&[\mubqperkg]  & [\mubqperkg]\\
\hline
copper detector support& 80~g/det. &$<0.088$     & $<16$         & $<19$  \\
PTFE  detector support & 10~g/det. &$0.60\pm0.11$& $ 25\pm 9$    & $ 31\pm14$\\
PTFE  insulation pipe  & 2.5~g/det.&$8\pm2$      &$1\,100\pm200$ & $<620$ \\
CC2 preamplifier       & per PCB~  &$1.8\pm0.3$  & $ 286\pm 28$  & $150\pm 24$\\
\end{tabular}
\end{center}
\end{table*}

 Finally, radioactivity measurements of the \gerda\ front-end electronics have
 been performed (see sec.~\ref{ssec:frontend}). Particular efforts have been
 made to produce a low radioactivity version of the circuit.  Some of the key
 points to achieve these results are: manufacturing of the printed circuit
 board on a specifically selected low-radioactivity substrate (Cuflon),
 minimization of the number of tantalum decoupling capacitors, integration of
 low value capacitors as stray capacitance between traces directly on board,
 and careful selection of passive physical components and soldering paste.  To
 reach a BI~$<$~\dctsper, the Monte Carlo predicts a maximum allowed activity
 for the front end electronics of 2~mBq for $^{226}$Ra and 500~\mubq\ for
 $^{228}$Th with a separation of 30~cm between the electronics and the top
 detectors. The average measured activity of a set of three preamplifiers is
 (286$\pm$28)~\mubq\ and (150$\pm$24)~\mubq\ in $^{226}$Ra and $^{228}$Th,
 respectively, including the pins.  Thus, the radiopurity limits are met for
 the 5~PCBs presently in use.

\section{Performance of the Apparatus}

 The construction of the apparatus was completed in June 2010. The
 commissioning phase started with the operation of refurbished
 \genat\ diodes from the GENIUS-TF experiment~\cite{geniustf}, in order to
 minimize the potential risks for the \geenr\ detectors. A larger
 background than expected at $Q_{\beta\beta}$ and an intense line at 1525~keV
 was discovered. The origin and mitigation was studied in the following months
 (see sec.~\ref{ssec:backgrounds}). In June 2011 a string of \geenr\ diodes
 was deployed for further preliminary tests including various operational
 configurations of the detectors and the electric stray fields. The
 commissioning phase was completed on November 1, 2011. All components had met
 their design specifications and an adequate background index was reached;
 thus, physics data taking of Phase I was started on November 9, 2011.  A
 blinding window of 40~keV width around \qbb\ is in place since January 11,
 2012. The raw data are written to disk, however events with energies from
 2019 to 2059~keV are not exported to the \tier1 data and are therefore not
 available for analysis. The blinding window will be opened when a sufficient
 exposure is acquired and the calibrations and selection cuts are finalized.

 While it is not the scope of this paper to discuss the physics analysis and
 results, the principal performance of the apparatus is summarized here.
 Results are shown demonstrating that a low background has actually been
 reached via thorough material selection and screening.  The stability of the
 performance of the complete \gerda\ setup at LNGS is inferred from the energy
 calibrations and the first spectra.  The performance particulars are obtained
 on the basis of physics runs between November 2011 and May 2012, which
 resulted in an exposure of 6.10~\kgy\ for the enriched detectors and
 3.17~\kgy\ for the natural detectors. These data are collected with an
 overall live time (calibration runs subtracted) of about 90\%.

\subsection{The performance of the muon veto}
   \label{sssec:muonperformance}

\begin{figure}[b]
\begin{center}
\ifmakefigures%
    \includegraphics[width=\columnwidth]{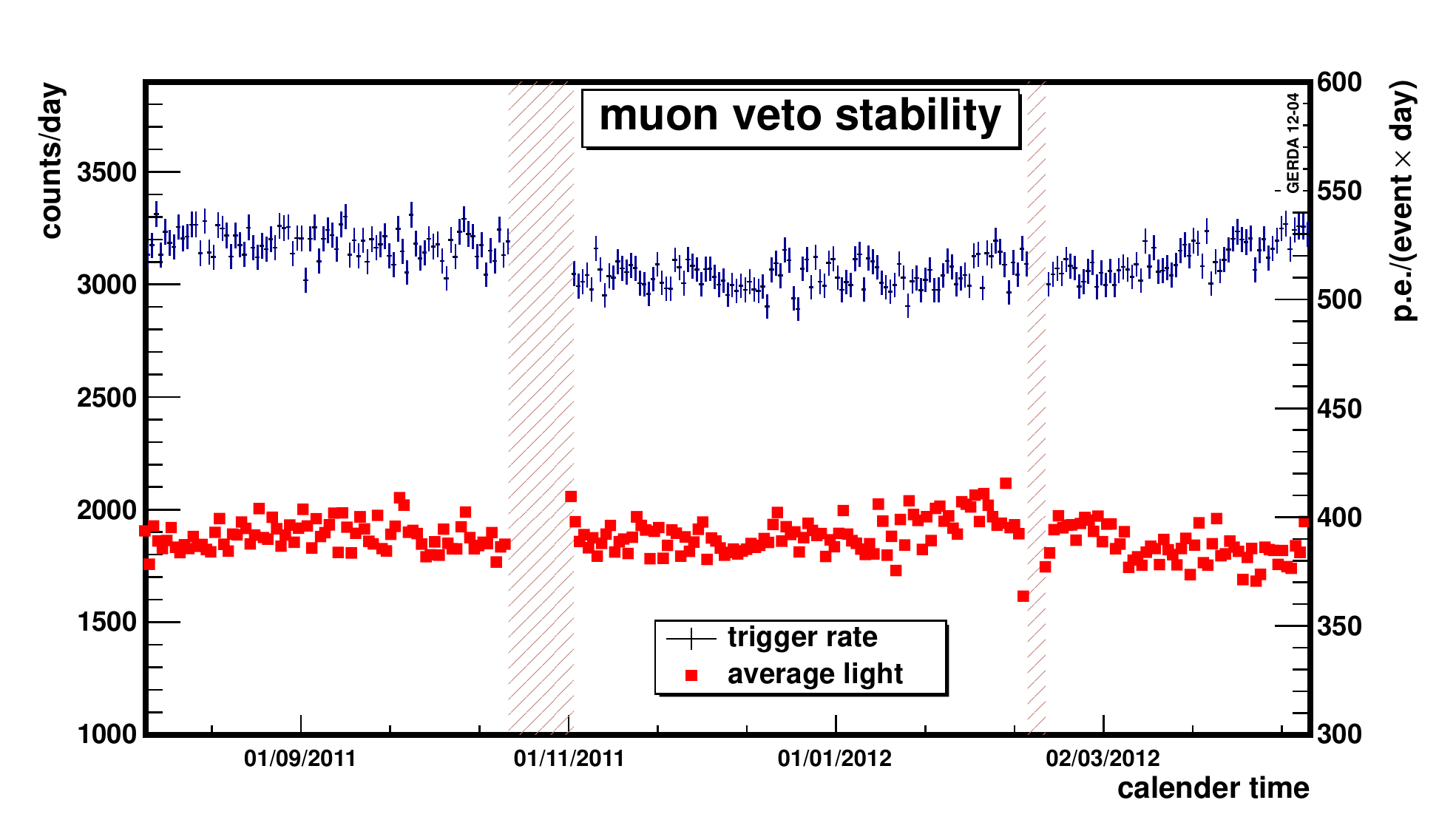}
\fi%
\caption{\label{fig:mvrate}
  The average light output per event and day (squares, right scale) of the
  Cherenkov muon veto. The daily rates (crosses, left scale) are rather
  constant.
}
\end{center}
\end{figure}

 The PMTs of the muon veto have been checked for pulse height stability for
 more than one year. A satisfying individual stability is reflected in the
 constant average light output per muon event per day (Fig.~\ref{fig:mvrate},
 squares and right scale). This constancy is mandatory for a reliable
 determination of the muon rate that is shown by the crosses in
 Fig.~\ref{fig:mvrate} (left scale). Except for short term fluctuation the
 rate is consistent with a 2~\% sinusoidal variation with a period of about
 one year. This is a well-known phenomenon~\cite{bellini12} that will be
 verified when a longer period of data is available.

 The observed muon rate in \gerda\ results in a preliminary value of
 (3.42$\pm$0.03)$\cdot$\powten{-4}~cts/(m$^2$s) which compares very well with
 the recent \borexino\ result of
 (3.41$\pm$0.01)$\cdot$\powten{-4}~cts/(m$^2$s) \cite{bellini12}.

\begin{figure}[b]
\begin{center}
\ifmakefigures%
      \includegraphics[width=\columnwidth]{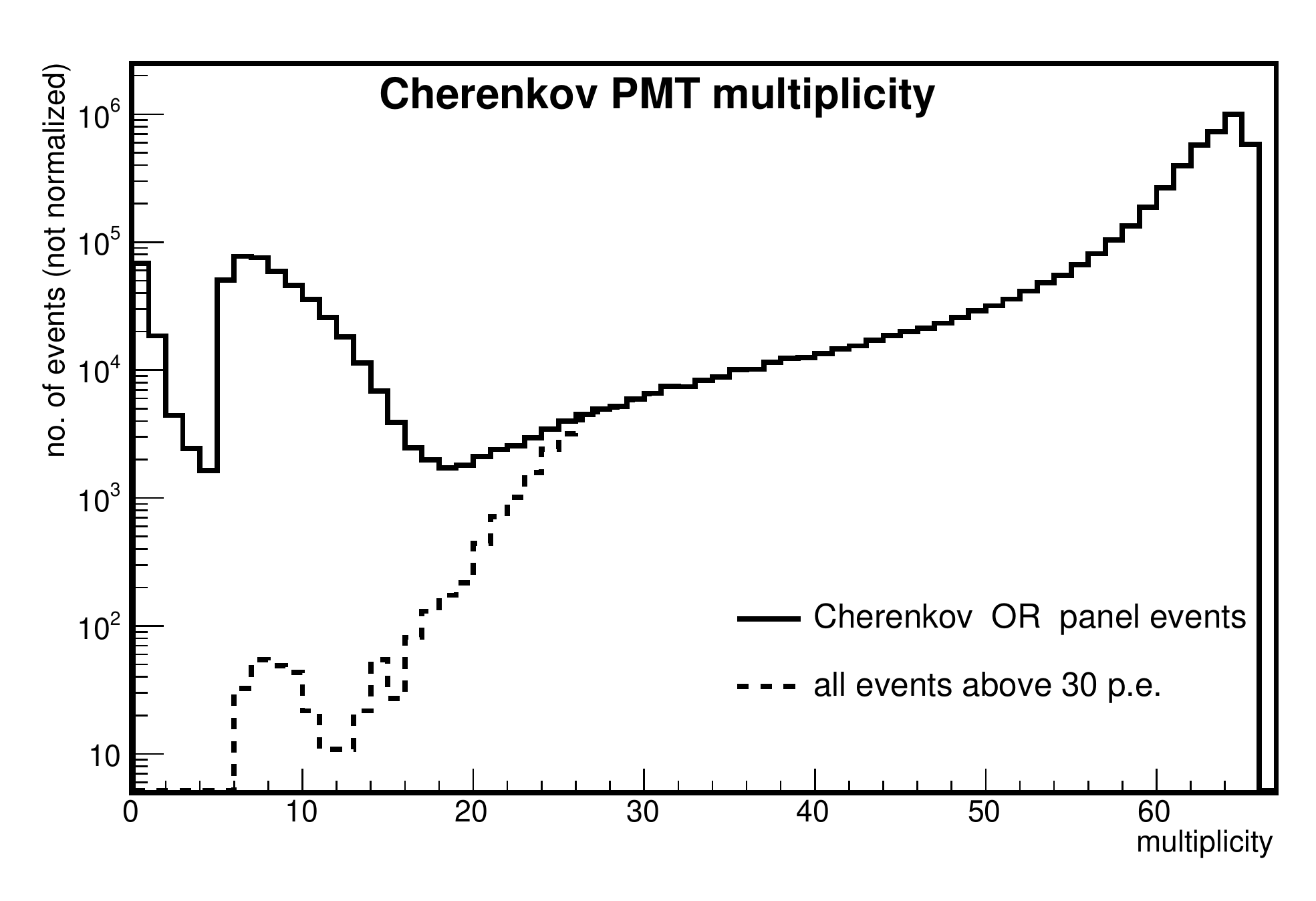}
\fi%
\caption{\label{fig:multiplicity}
  Measured multiplicities of all mounted Cherenkov PMTs without cut on the
  number of detected photoelectrons (full line) and with a cut of
  $\geq$30~p.e. (dashed line). 
}
\end{center}
\end{figure}

 Fig.~\ref{fig:multiplicity} shows the multiplicity~$M$, the number of
 Cherenkov PMTs fired. The spectrum is taken with trigger signals from both
 muon veto systems with a threshold of 1~photoelectron (p.e.).  The expected
 light yield is roughly 200 to 300 photons for every centimeter traversed by a
 muon. Since almost all surfaces of the water tank and cryostat are covered
 with the VM2000 foil, one would expect that most of the muon events
 will cause a high multiplicity of triggered PMTs.  The low coverage of 0.5\%
 of the surface by PMTs is compensated by the reflectivity and wavelength
 shifting properties of the VM2000 foil.  There is, indeed, a rise towards
 high multiplicities as predicted by the MC simulations, but there is also a
 prominent enhancement observed in the low multiplicity region below $M$=\,20
 which
 is not present in MC.  The low multiplicity bump around $M$=\,10 vanishes for
 events triggered by the plastic panels only.  Therefore, it is unlikely that
 it is caused directly by muons. The hypothesis of local radioactivity creating
 scintillation light in the VM2000 foil is still investigated. Triggers from
 the water Cherenkov cannot contribute to $M$$<$\,5.  The increase close to
 $M$=\,0 originates from triggers of the triple layered plastic scintillator
 when the muon hits the plastic but misses the water. Increasing the trigger
 threshold to 30~p.e. (dashed line) removes the intensity at low $M$.

 The lower limit of the muon detection efficiency (MDE) is estimated for a
 threshold of 30~{p.e.} amounting to
 $\epsilon_{md}$~=~(97.2$\pm$1.1)\,\%~\cite{phdRitter}.  MC
 estimates~\cite{phdKnapp} give a value of
 $\epsilon_{md}$~=~(99.1$\pm$0.4)\%. The latter, however, includes an energy
 deposition of the muon within the full detector array. This selects
 automatically longer tracks within the water, which in turn produce more
 detectable light.  A more refined determination employing coincidences
 between the plastic and Cherenkov detectors is under way.

\begin{figure*}[t]
\begin{center}
\ifmakefigures%
      \includegraphics[width=0.945\textwidth]{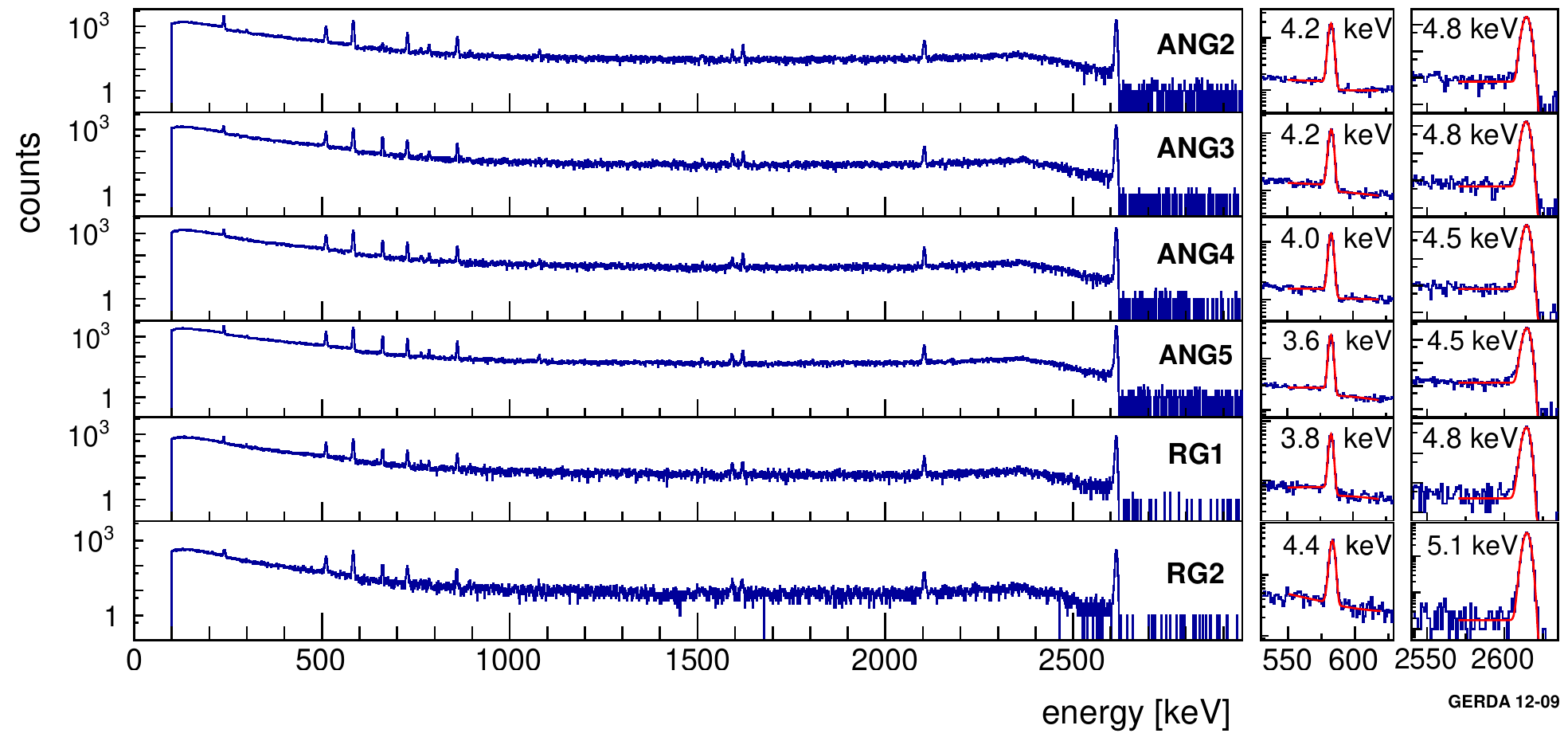}
\fi%
\caption{\label{fig:calibdets}
    The energy spectra of the six enriched germanium detectors are plotted for
    a calibration with \thzza. The blow-ups on the right show the fit results
    for the 583.2~keV and the 2614.5~keV lines including the values for a
    Gaussian FWHM.
}
\end{center}
\end{figure*}

 Alternatively, one can estimate the efficiency that a muon detected in the Ge
 detector array is accompanied by a muon trigger.  Two different event types
 were taken into account:
 (i) all events with {$>$\,8.5~MeV} deposited in germanium, but only {one}
     germanium  detector fired; and 
(ii) all events with {$>$\,4~MeV} deposited in germanium, but at least two
     germanium detectors  fired ($\alpha$ emitters  from the U/Th decay chain
     have energy  $>$\,4~MeV, but they would release their energy within one
     detector). 
 In the commissioning runs, a total of 79 events were selected by these cuts,
 while 78 of them were also found in the muon veto. The muon ``rejection
 efficiency'' is calculated as $\epsilon_{mr} = (97.9^{+1.2}_{-2.0})$~\%
 (median with 68~\% central interval), which is in a good
 agreement with the simulations.

 With the measured efficiency the background index due to un-identified prompt
 muons is estimated as $<10^{-5}$~\ctsper, which is well below the
 specifications needed for Phase~I and~II~\cite{phdKnapp}.

\subsection{Stability of Ge detectors}
 \label{ssec:stability}

 Initially, eight detectors from \igex\ and \hdm\ have been in operation in
 the \gerda\ cryostat.  Two of them, ANG~1 and RG~3, developed high leakage
 currents at the beginning of Phase~I. These detectors have been removed from
 the analysis of Phase~I data. For some time however, they have been used as
 veto to suppress multi-site events. The remaining total mass for analysis is
 14.6~kg with an average enrichment of 86\% in \gesix\ corresponding to
 165~moles.

\begin{figure}[b]
\begin{center}
\ifmakefigures%
       \includegraphics[width=\columnwidth]{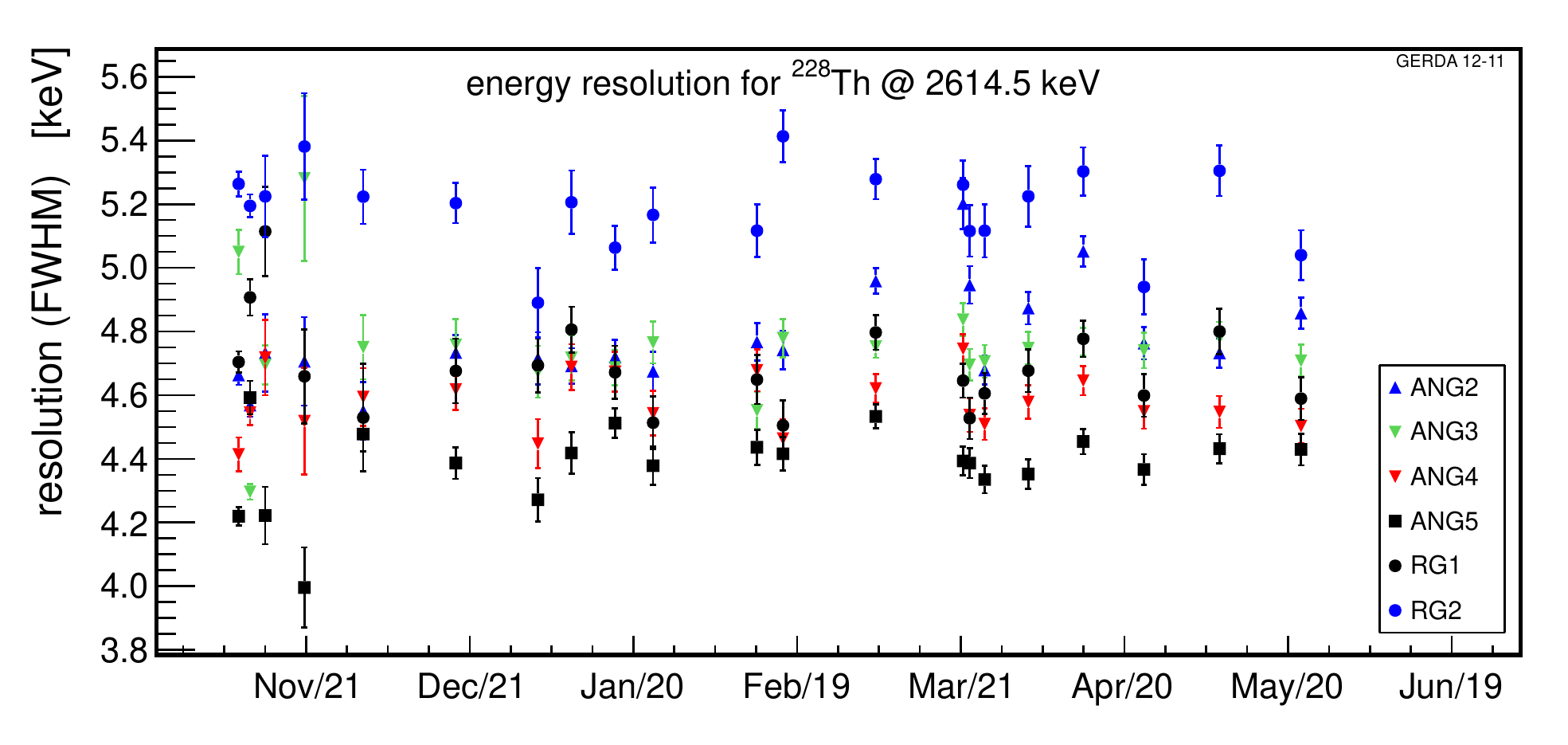}
\fi%
\caption{\label{fig:calibtime}
     The energy resolution of the germanium detectors is plotted for several
     energy calibrations with the \thzza\ source.
}
\end{center}
\end{figure}
 Energy calibrations are performed on a (bi)weekly basis with the
 \thzza\ sources.  Spectra of the six active enriched detectors are shown in
 Fig.~\ref{fig:calibdets}, including scaled subplots for the 583.2 and the
 2614.5~keV lines. The high count rates cause pile up that would manifest
 itself in tails on the low energy side of the peaks.
 Proper pile up rejection algorithms and further data quality cuts have been
 applied before the fitting~\cite{pipeline}.
\begin{table*}[t]
\begin{center}
\caption{ \label{tab:bi}
    The background index deduced (without pulse shape analysis) from the event
    count in the indicated energy windows $\Delta$E for different running
    conditions during the commissioning and the first part of Phase~I.
    Corresponding values are shown also for the \igex\ and
    \hdm\ experiments.
}
\vspace*{2mm}
\begin{tabular} {l c         c           c           c }
{ experiment}   & diodes   & $\Delta$E & exposure & background index\\
\hspace*{7mm}diode environment                    &         &    (keV)  & (kg$\cdot$yr) & 10$^{-2}$\ctsper \\
 \hline
{\igex} \cite{igex} \\   
\hspace*{7mm}vacuum, Cu enclosed            & enr & 2000-2500          &  4.7  & 26 \\
{\hdm} \cite{hdmabundances} \\   
\hspace*{7mm}vacuum, Cu enclosed            & enr & 2000-2100          & 56.7  & 16 \\
\hline
{\gerda\ commissioning}  \\
\hspace*{7mm}LAr                            & nat & 1839-2239          &  0.6  & 18$\pm$3 \\
\hspace*{7mm}LAr, Cu mini-shroud            & nat & 1839-2239          &  2.6  & 5.9$\pm$0.7 \\
\hspace*{7mm}ditto                          & enr & 1839-2239          &  0.7 & 4.3$_{-1.2}^{+1.4}$\\
{\gerda\ Phase~I}  \\
\hspace*{7mm}LAr, Cu mini-shroud             & nat & 1839-2239$^\star$ &  1.2 & 3.5$_{-0.9}^{+1.0}$\\
\hspace*{7mm}LAr (diodes AC-coupled)        & nat &  1839-2239$^\star$ & 1.9 & 6.0$_{-0.9}^{+1.0}$\\
\hspace*{7mm}LAr, Cu mini-shroud            & enr & 1939-2139$^\star$  &  6.1  & 2.0$_{-0.4}^{+0.6}$\\
\hline
\end{tabular} 
\end{center}
$\star$) excluding the blinded region of \qbb$\pm$20~keV 
\end{table*}
  The peaks are fitted well by a
 Gaussian and an error function representing the background. The results are
 shown by the red lines and the FWHM of the Gaussian is given in keV. Values
 between 4.2 to 5.3 keV (FWHM) at 2614~keV have been obtained. These can be
 translated to a mass weighted average of 4.5~keV (FWHM) at
 \qbb=2039.01(5)~keV~\cite{qvalue}.
 The resolution of the 2614.5~keV line for all detectors during the first
 months of data taking is shown in Fig.~\ref{fig:calibtime}.  No significant
 variation or trend is visible for this period.

\begin{figure}[h!]
\begin{center}
\ifmakefigures%
     \includegraphics[width=\columnwidth]{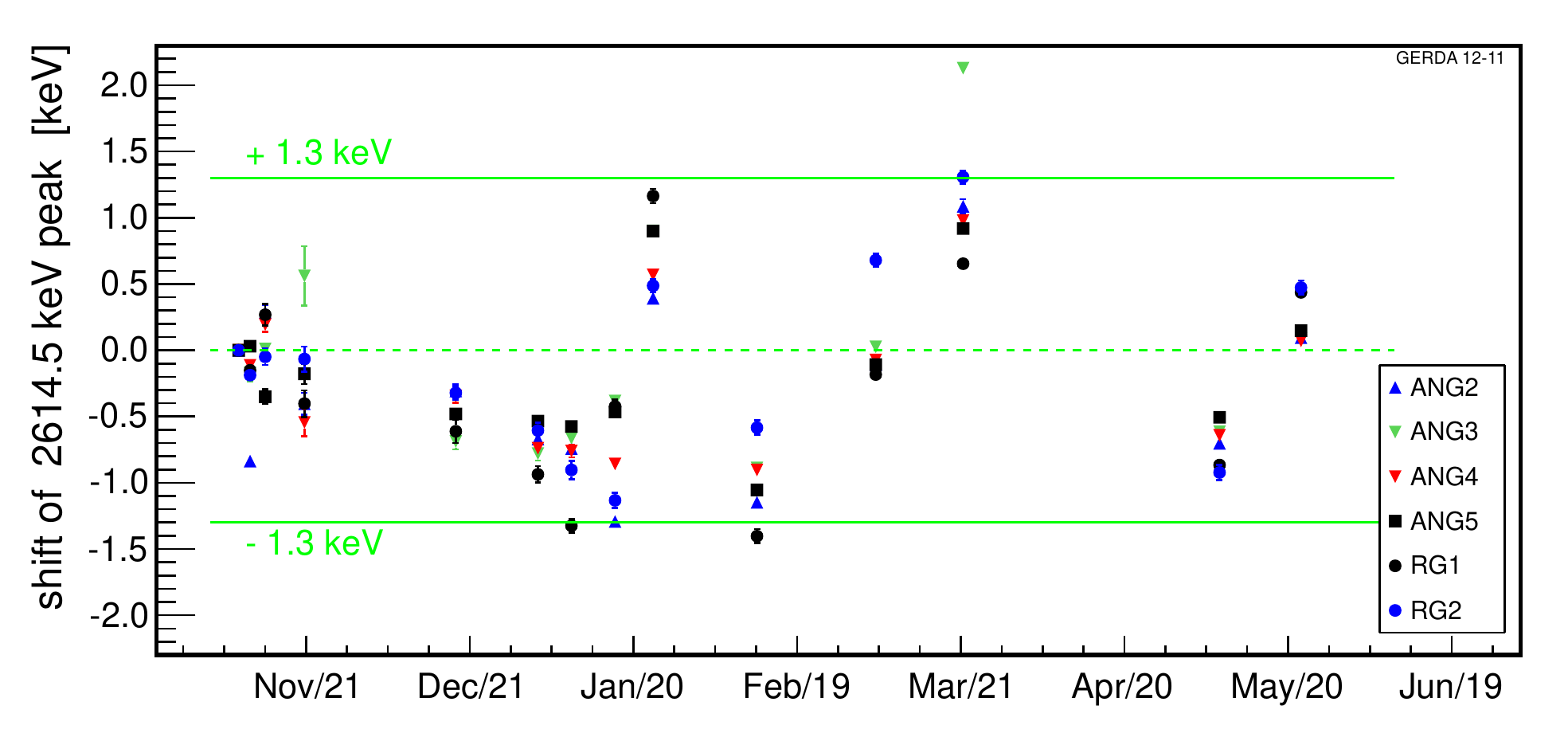}
\fi%
\caption{\label{fig:calstab}
   Variations of the 2614.5~keV $\gamma$ line between successive 
   calibrations. The green lines indicate  $\pm$1~keV variations at
   \qbb\ if scaled linearly in energy.
}
\end{center}
\end{figure}

 The same is also true for the gain, which normally changed only after some
 power cycling or temperature drifts.  The 2614.5~keV $\gamma$-line positions
 in the calibration spectra are stable in time as shown in
 Fig.~\ref{fig:calstab} as they fall into a range of $\pm$1.3~keV.  The shifts
 observed between two calibrations can be scaled linearly to the interesting
 energy at \qbb.  The two lines at $\pm$1.3~keV shown in
 Fig.~\ref{fig:calstab} correspond to $\pm$1~keV at \qbb.  The gain shifts
 within the ROI thus are typically less than 1~keV. This value is small
 compared to the average FWHM of 4.5~keV and shows that the data from all
 periods can be added in the search for the peak of the \onbb\ decay.

\subsection{Background levels in \textsc{Gerda}}
 \label{ssec:backgrounds}

 The commissioning of \gerda\ started with a string of three bare low
 background \genat\ diodes, and yielded a surprisingly large BI on the order
 of the \hdm\ and \igex\ experiments ($18\cdot10^{-2}$~\ctsper, see
 Table~\ref{tab:bi}).  As another surprise, the line at 1525~keV from
 $^{42}$K, the daughter of $^{42}$Ar, appeared in the spectra with an
 intensity much higher than expected on the basis of the upper limit for the
 ratio $^{42}$Ar/$^{nat}$Ar determined by V.D. Ashitkov \etal~\cite{ar42}.
 The published limit of $<$4.3$\times$10$^{-21}$~g/g at 90\% confidence level
 converts to an upper limit of 41~\mubq/kg for $^{42}$Ar.  These observations
 led to the working hypothesis that charged ions, and in particular the
 progeny $^{42}$K, are drifting in the electric field of the bare Ge diodes
 that are biased with 3 to 4~kV via the large n$^{+}$ surface (see
 sec.~\ref{sec:det} and Fig.~\ref{fig:detschem}). The concentration of
 radioactive impurities near the Ge diodes can increase.  Further studies with
 different bias schemes confirmed this hypothesis. A major improvement of the
 BI was achieved by enclosing the string of detectors with a cylinder, made of
 60~\mum\ thick Cu foil, called ``mini-shroud'' (BI~$\approx 5.9\cdot
 10^{-2}$~\ctsper).  The volume of LAr from which the ions can be collected
 onto the surface of the detectors is reduced and bulk convection of the LAr
 near the detectors is prevented.  In fact, operating the Ge diodes in
 AC-coupled mode ($n^+$ surface grounded and $p^+$ contact biased) without
 mini-shroud but with adequate shielding of the $p^+$ contact, i.e. without
 external electrical stray field, yielded a similar BI of
 $6.0\cdot10^{-2}$~\ctsper\ (see next to last line in Table~\ref{tab:bi}).
 For the Phase~I physics run, the hermeticity of the mini-shroud, as well as
 the shielding of the HV cables, was further improved in order to avoid any
 leakage of electric field lines into the LAr volume. The improvement with
 respect to the precursor experiments is evident.  The stability of the BI
 must be proven for a longer period of time.

\begin{table*}[t]
\begin{center}
\caption{   \label{tab:backgroundLines}
    Counts and rates of background lines for the enriched and natural
    detectors in \gerda\ in comparison to the enriched detectors of
    \hdm~\cite{oleg}.  Upper limits correspond to $90~\%$~credibility
    interval. The central value is the mode of the posterior probability
    distribution function and the error bars account for the smallest interval
    containing 68\% probability.
}
\vspace*{1mm}
   \begin{tabular}{c c |r@{ / }lc| r@{ / }lc| r@{ $\pm$ }l }
      &
      & \multicolumn{3}{c|}{\genat\ (3.17\,\kgyr)}
      & \multicolumn{3}{c|}{\geenr\ (6.10\,\kgyr)}
      & \multicolumn{2}{c}{\hdm\ (71.7\,\kgyr)} \\
      isotope & energy 
      & \multicolumn{2}{c}{tot/bck} & rate
      & \multicolumn{2}{c}{tot/bck} & rate
      & \multicolumn{2}{c}{rate}\\
      & \small[keV] 
      & \multicolumn{2}{c}{\small[cts]} & \small[cts/(kg$\cdot$yr)]
      & \multicolumn{2}{c}{\small[cts]} & \small[cts/(kg$\cdot$yr)]
      & \multicolumn{2}{c}{\small[cts/(kg$\cdot$yr)]}\\
      \hline
\up
      $^{40}$K   & 1460.8 &  85 &  15 & $21.7_{- 3.0}^{+ 3.4}$  &   125 &    42 & $13.5_{- 2.1}^{+ 2.2}$  &  181 & 2   \\  
\up
      $^{60}$Co  & 1173.2 &  43 &  38 & $<5.8$                  &   182 &   152 & $ 4.8_{- 2.8}^{+ 2.8}$  &   55 & 1   \\  
                 & 1332.3 &  31 &  33 & $<3.8$                  &    93 &   101 & $<3.1$                  &   51 & 1   \\  
      $^{137}$Cs & 661.6  &  46 &  62 & $<3.2$                  &   335 &   348 & $<5.9$                  &  282 & 2   \\  
      $^{228}$Ac & 910.8  &  54 &  38 & $ 5.1_{- 2.9}^{+ 2.8}$  &   294 &   303 & $<5.8$                  & 29.8 & 1.6 \\  
\up
                 & 968.9  &  64 &  42 & $ 6.9_{- 3.2}^{+ 3.2}$  &   247 &   230 & $ 2.7_{- 2.5}^{+ 2.8}$  & 17.6 & 1.1 \\  
      $^{208}$Tl & 583.2  &  56 &  51 & $<6.5$                  &   333 &   327 & $<7.6$                  &   36 & 3   \\[0.15cm]
                 & 2614.5 &   9 &   2 & $ 2.1_{- 1.1}^{+ 1.1}$  &    10 &     0 & $ 1.5_{- 0.5}^{+ 0.6}$  & 16.5 & 0.5 \\  
\up
      $^{214}$Pb & 352    & 740 & 630 & $34.1_{-11.0}^{+12.4}$  &  1770 &  1688 & $12.5_{- 7.7}^{+ 9.5}$  & 138.7 & 4.8 \\  
\up
      $^{214}$Bi & 609.3  &  99 &  51 & $15.1_{- 3.9}^{+ 3.9}$  &   351 &   311 & $ 6.8_{- 4.1}^{+ 3.7}$  &  105 & 1   \\  
\up
                 & 1120.3 &  71 &  44 & $ 8.4_{- 3.3}^{+ 3.5}$  &   194 &   186 & $<6.1$                  &  26.9 & 1.2\\  
\up
                 & 1764.5 &  23 &   5 & $ 5.4_{- 1.5}^{+ 1.9}$  &    24 &     1 & $ 3.6_{- 0.8}^{+ 0.9}$  & 30.7 & 0.7 \\  
\up
                 & 2204.2 &   5 &   2 & $ 0.8_{- 0.7}^{+ 0.8}$  &     6 &     3 & $ 0.4_{- 0.4}^{+ 0.4}$  &  8.1 & 0.5 \\
\end{tabular}
\end{center}
\end{table*}

 An analysis of the intensity of the 1525~keV line gives a concentration for
 $^{42}$Ar that is about twice the literature limit.  This estimate is based
 on the assumption of a homogeneous distribution of this isotope outside the
 mini-shroud.

 The intensity of $\gamma$ lines was investigated in order to identify sources
 of backgrounds. The results are compiled in Table~\ref{tab:backgroundLines}
 for the natural and the enriched detectors in comparison to numbers from
 \hdm~\cite{oleg}.  The rate estimates are based on a Bayesian approach
 starting with a flat prior probability distribution function.  The general
 observation is an achieved reduction by about a factor of 10 with respect to
 the \hdm\ experiment.  The composition of the background in relation to the
 screening results will be discussed in a future publication.

 Additional contributions to the BI will result from radioactive surface
 contaminations such as $^{210}$Pb as well as from cosmogenically produced
 radioisotopes within the diodes.  These contributions are expected to be
 small and will require large data sets to evaluate.

 Finally, it is worth to mention that auxiliary experiments were performed to
 study the cross sections of cosmogenic activation of steel and other
 constructional materials~\cite{laubenheusser}, the inelastic neutron
 scattering~\cite{domula}, the neutron activation cross sections, and the
 $\gamma$ decay spectra~\cite{georg,marga,georgc}. The deduced contributions
 to the BI are in the order of few 10$^{-5}$~\ctsper.
 
\begin{figure}[b]
\begin{center}
\ifmakefigures%
        \includegraphics[width=\columnwidth]{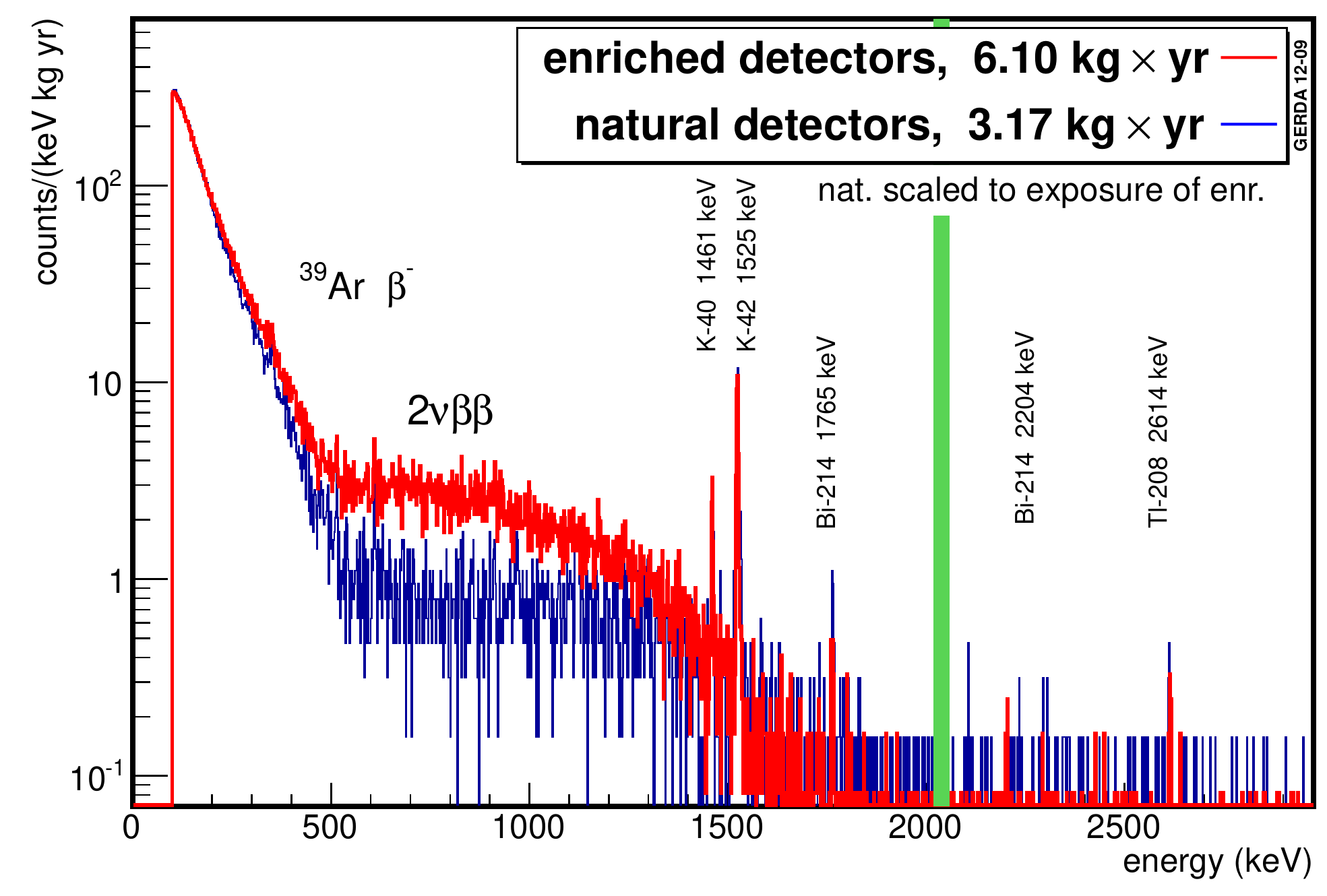}
\fi%
\caption{\label{fig:spectra-ge}
     Spectra taken with enriched (red) and non-enriched (blue) detectors
     during the same time period. The \genat\ spectrum has been normalized to
     match the exposure of \geenr.  The blinding window of \qbb$\pm$20~keV is
     indicated as green bar. Identified $\gamma$ lines are indicated.
}
\end{center}
\end{figure}

 First energy spectra for enriched and natural diodes are shown in
 Fig.~\ref{fig:spectra-ge}.
 Notice, that the spectrum from the natural detectors has been renormalized to
 match the exposure of the enriched diodes.  The low energy part is dominated
 by the $\beta$ decay of $^{39}$Ar which has an endpoint energy of 565~keV.
 The well known activity of $A(^{39}$Ar)=
 [1.01$\pm$0.02(stat)$\pm$0.08(syst)]~Bq/kg (Ref.~\cite{benetti39Ar})
 describes the observed intensity.  The enhancement of \nnbb\ events in the
 range from 600 to 1400~keV for the enriched detectors is clearly visible.

 The BI of $(2.0_{-0.4}^{+0.6})\cdot$~\zctsper\ for the enriched detectors is
 evaluated in the energy region \qbb$\pm$100~keV with the 2019 to 2059~keV
 window excluded (green bar in Fig.~\ref{fig:spectra-ge}).  This value is an
 order of magnitude lower than the one for the very same detectors in their
 previous shielding in the \hdm\ and \igex\ experiments (see
 Table~\ref{tab:bi}).
\section{Conclusions}
\label{sec:conclusions}

 \gerda\ searches for \onbb\ decay of $^{76}$Ge using a new experimental
 concept.  Bare germanium diodes are operated successfully in a 4~m diameter
 cryostat filled with LAr, requiring only a small amount of radiopure
 materials as mechanical and electrical support.  Shielding against external
 background is achieved by LAr and an additional shell of 3~m of water.

 The experiment started commissioning in May 2010 and in November 2011
 with physics data taking (Phase~I). The experience gained so far shows
 that all components work well.
\begin{enumerate}
 \item  The operation of the cryostat inside the water tank  is stable and safe.
 \item Bare germanium diodes are operated reliably in liquid
       argon over a long time and the implemented handling procedure
       ensures that many operational cycles do not deteriorate the
       performance.
 \item The readout electronics is balancing the partially
      conflicting requirements of good energy resolution, low radioactivity,
      and operation at LAr temperature.
 \item The water tank instrumentation ensures a high veto efficiency of
       muon events and only a tolerable loss of 3 out of 66
       PMTs have stopped functioning during a 2~year period. 
 \item Data acquisition and monitoring of the ambient parameters operate
   reliably. 
 \item The implemented software allows for a fast reconstruction of the
       data together with a good monitoring of data quality.
\end{enumerate}

 The experience from the (bi-)weekly calibrations shows that the gain drifts
 of the entire readout chain are typically smaller than 1~keV at
 $Q_{\beta\beta}$. This is small enough to ensure that adding all data will
 not result in relevant shifts of peak positions or deteriorations of
 resolutions.

 The surprisingly large background from $^{42}$K, the $^{42}$Ar progeny,
 experienced during the commissioning can be mitigated by two methods:
 encapsulation of each detector string by a closed thin-walled copper cylinder
 or AC coupling of the detectors. In both cases the electric field outside of
 the encapsulation is minimized.
 
 The Phase~I background is determined currently to
 $(2.0_{-0.4}^{+0.6})\cdot$\zctsper. This value and the intensities of gamma
 lines show an order of magnitude improvement compared to the previous
 \hdm\ and \igex\ experiments. In the absence of a signal and given the
 current BI, \GERDA\ expects to set 90~\% probability lower limits of
 $T_{1/2}>1.9\cdot10^{25}$ yr for an exposure of 20~\kgyr.

\section*{Acknowledgments}
 The \gerda\ experiment is supported financially by
   the German Federal Ministry for Education and Research (BMBF),
   the German Research Foundation (DFG) via the Excellence Cluster Universe,
   the Italian Istituto Nazionale di Fisica Nucleare (INFN),
   the Max Planck Society (MPG),
   the Polish National Science Centre (NCN),
   the Russian Foundation for Basic Research (RFBR), and
   the Swiss National Science Foundation (SNF).
 The institutions acknowledge also internal financial support.

The \gerda\ collaboration thanks the directors and the staff of the LNGS
for their continuous strong support of the \gerda\ experiment.

Preparing and setting up  the infrastructure of the  \gerda\ experiment
was made possible only through the indispensable help of 
 R.~Adinolfi Falcone,
 T.~Apfel,
 P.~Aprili,
 J.~Baumgart,
 G.~Bucciarelli,
 M.~Castagna,
 F.~Costa,
 N.~D'Ambrosio,
 S.~Flicker,
 D.~Franciotti,
 H. Fuchs,
 H.~Hess,
 V.~Mallinger,
 P.~Martella,
 B.~M{\"o}rk,
 D.~Orlandi,
 M.~Reissfelder,
 T.~Schwab,
 R.~Sedlmeyr,
 S.~Stallio,
 R.~Tartaglia,
 E.~Tatananni,
 M.~Tobia,
 D. Wamsler,
 G.~Winkelm{\"u}ller, and  
 T.~Weber.

The prolific cooperation with P.~Vermeulen and J. Verplancke from Canberra
SNV, Olen in the context of refurbishment of the enriched germanium diodes is
appreciated.




\begin{thebibliography}{000}
\bibitem{makrolon}
Makrolon is a registered trade mark of BASF.
\bibitem{y11}
  S-type Y11(250)M, Kuraray, Japan.  (The middle of the fiber is made of
  polystyrol with additive 0.02\% of wavelength shifter Y11.  The internal
  part of fiber consists of PMMA, external from the fluorinated PMMA).
\end{thebibliography}
\end{document}